# Electric Vehicle Charging: a Survey on the Security Issues and Challenges of the Open Charge Point Protocol (OCPP)

Zacharenia Garofalaki, Dimitrios Kosmanos, Sotiris Moschoyiannis, Dimitrios Kallergis and Christos Douligeris

***Abstract*—**The increased use of smart Electric Vehicles (EVs) and Plug-in Electric Vehicles (PEV) opened a new area of research and development. The number of EV charging sites has considerably increased in residential as well as in public areas. Within these EV charging sites, various entities need to communicate in a secure and efficient way. The Open Charge Point Protocol (OCPP) offers a way to coordinate this communication and is already being used in many implementations. However, only the latest OCPP 2.0 version of the protocol includes certain security features. In this article, we present the entities that take part in an OCPP-based smart charging scenario, we identify security issues and threats and present solutions that have been proposed by scholars. We identify open security issues for OCPP and propose future research directions for the security enhancement of the protocol.

## I. Introduction

The wide adoption of smart electric vehicles has opened a new area of research and development. It is predicted that by 2040, one out of three cars globally will be electric. Some of the biggest automobile manufacturers worldwide are planning on introducing several new electric vehicle models each, in the next 10 years [1]. This turn of the global car fleet towards Electric Vehicles (EVs) demands for an expansion and enhancement of the existing EV charging infrastructure.

Open standards and a shared infrastructure for the EV charging are vital for the unification of the service provided by independently operating charging stations and charging lots. This unification will allow EVs to gain access to various cross-vendor charging stations, charge frequently and, therefore, operate seamlessly. This approach, however, necessitates the coordination of the services, the charging station operations, and the EVs among others. EVs are part of and will be integrated with the smart transportation and smart electric infrastructure that are connected to, forming a complex system consisting of a variety of entities and technologies [2]. This new system that arises from the interconnection of smart cars with the smart grid consists of mobile devices, autonomous cars, and heterogeneous cyber-physical systems, exposing all the above to new threats and vulnerabilities [3]. Even though security technologies have already been integrated in some systems [4], there is a need for these technologies to be adapted for addressing the specific challenges of the EV charging infrastructure.

Since several entities need to communicate in a secure and efficient way in an EV charging system, also referred to as Plug-in Electric Vehicle (PEV) network, a plethora of protocols and standards are used to regulate this network communication. Organisations such as the International Organization for Standardization (ISO), the International Electromechanical Commission (IEC), the Society of Automotive Engineers (SAE), and the Institute of Electrical and Electronics Engineers (IEEE), contribute in this standardisation effort. However, the Open Charge Point Protocol (OCPP) stands out as the de facto used protocol in 148 countries across all 6 continents and is supported by more than 65,000 installed and operating charging stations [5]. It is also reported that more than 40 electric chargers vendors are integrating OCPP in their products [6],[7]. The protocol is supported by the Open Charge Alliance (OCA) with more than 220 member-companies active in the electric mobility area [8]. OCPP [9] is involved in the reservations and the management of the charging processes to ensure quality of service (QoS) and efficiency in terms of charging. The main advantages that led to the OCPP prevalence over other protocols are that it is an open and free protocol, and it supports cross-vendor operability, as well as fast and easy integration of devices.

The security of the EV charging systems is interdependent to the security of the EVs, of the drivers, the infrastructure of the system itself, the power supplier, and the power grid. As electric mobility is getting more popular, security attacks against these assets are more common. In the period 2019-2021, cyber-attacks against EVs increased by 225% [10]. A study revealed severe vulnerabilities in the charging stations of 16 different vendors that could be exploited [11], while Acharya et al. [12] mentioned that local power supplies are vulnerable to the lack of security on EVs and EV chargers data.

The current survey is expanding the existing research literature regarding the security of the EV charging systems to include the security and privacy issues of all the assets of an EV charging system. More than that, the current survey is focused on the latest of the four released versions of the protocol, the OCPP version 2.0.1 [13]. This version, also referred to as OCPP 2.0, was released in 2018 and includes improved security features. OCPP 2.0 is the improvement of the version 1.6 [14], which introduced limited security features for the protocol and was released in 2015. The latest OCPP 2.0 version incorporates features, such as secure firmware updating, logging and event notification, and secure authentication [15], as well as support for Transport Layer Security (TLS) and key management of the client-side certificates. The recent addition of security features on the protocol indicates that the study of the protocol's level of security is ongoing and of high importance.

The contributions of the current survey are:

*(a)* the depiction of a typical architecture and the description of the entities that take part in an OCPP-based EV charging system;
*(b)* an analysis of the security issues and threats, and of the related countermeasures proposed by other scholars;
*(c)* an association of the aforementioned security threats with every asset of an EV charging system; and
*(d)* the identification of open issues and a proposal regarding future research directions.

The rest of this article is structured as follows. In Section II, the architecture of a typical OCPP-based EV charging system is depicted. The elements within the architecture are described with emphasis on the OCPP protocol's history and characteristics. The other standards and protocols that co-exist with OCPP within a PEV network and cover end-to-end communication are presented as well. In Section III, other recent surveys in the area of EV charging systems security are shown and compared against this work. In Section IV, the security issues that arise from the interconnection of smart cars with the smart grid and the existing solutions are presented. In Section V, the privacy issues, and the detection and deflection mechanisms are discussed. Based on the aforementioned, in Section VI we identify open issues in security and privacy of an OCPP-based EV charging system, and we propose future research directions. Finally, our conclusions are presented in Section VII. A list of the acronyms used throughout this paper is shown in Table I.

TABLE I
ACRONYMS

| Acronyms | Description | Acronyms | Description |
|---|---|---|---|
| AC | Alternating Current | MIB | Management Information Base |
| AEC | Automotive Electronics Council | MitM | Man-in-the-Middle |
| AES | Advanced Encryption Standard | NFC | Near-field Communication |
| AI | Artificial Intelligence | OCA | Open Charge Alliance |
| AMI | Advanced Metering Infrastructure | OCHP | Open Clearing House Protocol |
| ARM | Architectural Reference Model | OCPI | Open Charge Point Interface |
| ARP | Address Resolution Protocol | OCPP | Open Charge Point Protocol |
| BC | Blockchain | OEM | Original Equipment Manufacturer |
| BDD | Binary Decision Diagrams | OFDM | Orthogonal Frequency Division Multiplexing |
| BPNN | Back Propagation Neural Network | OICP | Open InterChange Protocol |
| CA | Certificate Authorities | OMS | Outage Management System |
| CAN | Controller Area Network | OpenADR | Open Automated Demand Response |
| CC | Common Criteria | OSCP | Open Smart Charging Protocol |
| CCert | Contract Certificate | OTA | Over-The-Air |
| CHAdeMO | CHArge de MOve | P2P | Peer-to-Peer |
| CPO | Charging Point Operator | PBFT | Practical Byzantine Fault Tolerance |
| CPS | Cyber Physical System | PEV | Plug-in Electric Vehicle |
| CS | Charging Station | PHY | Physical Layer |
| CSMS | Charging Station Management System | PK | Private Key |
| DC | Direct Current | PKI | Public Key Infrastructure |
| DDoS | Distributed Denial of Service | PLC | Power Line Communication |
| DER | Distributed Energy Resources | PnC | Plug & Charge |
| DLT | Distributed Ledger Technology | PoS | Proof of Stake |
| DOE | Department of Energy | PoW | Proof of Work |
| DoS | Denial of Service | PUF | Physical Unloneable Functions |
| DSO | Distribution System Operator | PWM | Pulse Width Modulation |
| EAL | Evaluation Assurance Level | QoS | Quality of Service |
| ECC | Elliptic Curve Cryptography | RBAC | Role-Based Access Control |
| ECU | Electric Control Unit | RCE | Remote Control Execution |
| EMS | Energy Management System | RFID | Radio Frequency IDentification |
| EMSP | E-Mobility Service Provider | RKE | Remote Keyless Entry |
| EMV | Europay, Mastercard, and Visa standard | RSU | Restricted Stock Unit |
| ETDS | Energy Theft Detection System | SAE | Society of Automotive Engineers |
| EV | Electric Vehicle | SC | Smart Card |
| EVCC | Electric VehicleCommunication Controller | SDN | Software Defined Network |
| EVSE | EV Supply Equipment | SDP | SECC Discovery Protocol |
| EXI | Efficient XML Interchange | SECC | Supply Equipment Communication Controller |
| FAN | Field Area Network | SEP | Smart Energy Profile |
| FDIA | False Data Injection Attack | SG | Smart Grid |
| FPR | False Positive Rate | SK | Secret Key |
| GPS | Global Positioning System | SNMP | Simple Network Management Protocol |
| HSM | Hardware Security Module | SOAP | Simple Object Access Protocol |
| HTTP | Hypertext Transfer Protocol | SOC | State Of Charge |
| IDS | Intrusion Detection System | SoC | System on Chip |
| IEC | International Electrotechnical Commission | SSL | Secure Sockets Layer |
| IED | Intelligent Electronics Device | TCP | Transmission Control Protocol |
| IEEE | Institute of Electrical and Electronics Engineers | TLS | Transport Layer Security |
| IP | Internet Protocol | TPM | Trusted Platform Module |
| ISO | International Organization for Standardization | TSO | Transmission System Operator |
| IT | Information Technology | UDP | User Datagram Protocol |
| JA | Judging Authority | USB | Universal Serial Bus |
| JSON | JavaScript Object Notation | V2G | Vehicle-to-Grid |
| KH | Krill Herd algorithm | V2GTP | V2G Transport Protocol |
| LAG | Link Aggregation Group | V2H | Vehicle-to-Home |
| LAN | Local Area Network | V2X | Vehicle-to-everything |
| LC | Local Controller | VAS | Value Added Service |
| LP | Local Proxy | WAN | Wide Area Network |
| LS | Least Squares | WPT | Wireless Power Transfer |
| MAC | Message Authentication Code | XML | Extensible Markup Language |
| MAC address | Media Access Control address | | |

## II. OVERVIEW

OCPP was the result of an idea proposed by the Dutch foundation ElaadNL for an open protocol to support the communication between the charging points and the back-end systems. The foundation's recommendation was that OCPP should be standardized and certified by OCA. This OCA certification would ensure uniformity and cross-vendor compatibility [16]. Since 2014, OCPP has been owned by OCA [17].

OCPP is a demand-response protocol which mainly provides the messages for the communication between the Charging Station (CS) and the Charging Station Management System (CSMS), although in practice it is not exclusively used for that communication. In this Section, we present an Architectural Reference Model (ARM) of the topology of a PEV charging network supported by OCPP, a brief description of the main elements in the ARM, the OCPP-based communication model, the power trading management process, and the protocols and standards that coexist in a PEV network.

The architecture and the elements of an OCPP-based EV charging system, the communication model and the power trading management supported by OCPP, as well as the co-existing protocols and standards within an EV charging system, are also presented.

### A. Architecture of the EV charging system

The architecture of an OCPP-based EV charging system, as shown in Figure 1, contains the main entities that co-operate in the lifecycle of the charging service. The architecture depicts *(a)* all the devices that are located on the charging site, including the consuming EV in a service's instance, *(b)* the CSMS and the Distribution System Operator (DSO), and *(c)* the protocols or standards applied on the component-to-component connections. This architecture is based on the topologies that are supported by OCPP, and which were introduced by OCA [18].

Every EV charging system implementation may embed some or all of the components described in the architecture. The minimal implementation, however, requires the *main components*, namely a CSMS, at least one CS, and the Electric Vehicle Supply Equipment (EVSE) for the vehicle's plugging and charging.

The existence of a direct communication between certain components is mandatory. In some cases, the communication between two components may or may not be explicit. For instance, if the implementation includes a DSO, i.e. a third-party component, this participation is realized by the direct communication between the DSO and the Energy Management System (EMS). A communication between the DSO and the CSMS may also exist, if this facilitates both the EV charging system and the third party.

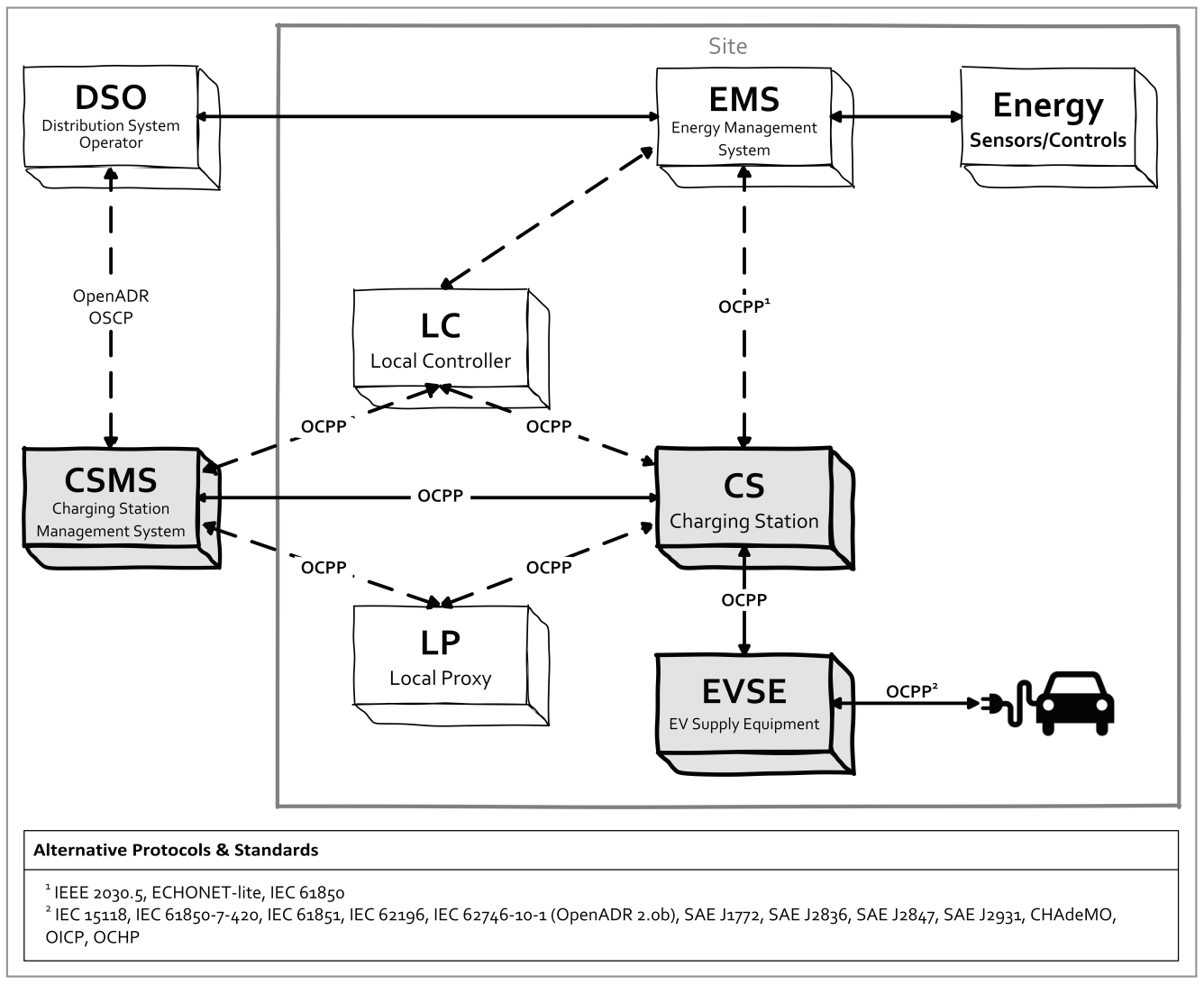

Fig. 1. Architecture of an OCPP-based EV charging system

The list of protocols and standards included in the architecture is restricted to the ones that define the communication of the *main components*, hence to the ones that define the EV charging system operation. OCPP mainly supports the communication amongst the three *main components*. This means that the CS and the EVSE have an OCPP-based communication, and the same happens between the CSMS and the CS when they are directly connected, and when a Local Controller (LC) or a Local Proxy (LP) mediates. Moreover, OCPP supports the communication between *(a)* the CS and the EMS, and *(b)* the EV and the EVSE [19].

It should be noted that OCPP is a demand-response protocol used by IP-connected devices which communicate via the TCP/IP protocols stack. Therefore, the protocol's operations and security are dependent on the components of the architecture, their operational characteristics, and their connectivity.

### B. Elements of the charging infrastructure

The components of an EV charging system architecture are the following.

*1) Electric Vehicle Supply Equipment (EVSE):* EVSE is the core subsystem of a CS providing the interface for the consuming EV to connect and charge. EVSE is the edge system of the charging system which collects the EV data regarding the EV's charging and connectivity status. OCPP 2.0, which refers to the 2.0.1 version of the protocol, supports the data exchange between the EVSE and the EV upon connection [20] and provides a set of standard messages for the communication between the EVSE and the CSMS [21].

*2) Charging Station (CS):* CS is the managing system of one or of a group of charging points and is located in a short-range area of the charging site. The charging point hosts the EVSE and the EV's power connectors. CS is controlled by the CSMS; CSMS creates the messages that define the power limits and the operational state of the CS. Moreover, CS controls the charging process of the connected EVs and enforces these limits. OCPP 2.0 supports the authentication, transaction and billing requirements, leading to the specification of the per case limits [22],[23].

*3) Local Controller (LC):* LC is an optional controller for the control of one or several CSs. LC either intervenes and facilitates the CS-CSMS communication or controls the charge limits on the CS when the CS-CSMS communication is lost or interrupted. The use of scattered LCs in a PEV network may be implemented to support and backup the CSMS, and to distribute the control procedures and their computational cost. OCPP includes functions for the use cases that include LCs. These functions are built on the assumption that an LC is a CS without any EVSE or any connector [18].

*4) Local Proxy (LP):* LP is an optional unit which acts as a router. The LP is used to route messages from and to one or more CSs, especially if the CS has no access to the network, due to their positioning, e.g. they are located underground. In such a topology, the OCPP structure dictates the CS to communicate with the LP as if the latter is the CSMS, and vise versa.

*5) Charging Station Management System (CSMS):* CSMS is the coordinator of an EV charging system. CSMS's main tasks [19] are the following: *(a)* to communicate with the CS and the EVSE, *(b)* to define the service's parameters taking into account the user's input, and the EV's and the power grid's status, *(c)* to collect and store the charging system's data, *(d)* to host the user application, and *(e)* to maintain a booking registry for the service. CSMS communicates with the charging system's components via OCPP, with the exception of the communication between the CSMS and the DSO. OCPP supports smart-charging policies and allows the CSMS to implement customized profiles for the charging processes [21].

*6) Distribution System Operator (DSO):* DSO is the system or organization responsible for distributing electricity to the end-users. The DSO allows or prohibits the power flow to the charging site, and, based on the EV's data feedback, ensures balance and decongestion in the grid [24]. At least one edge system of the operator is considered a third-party component in the EV charging system. This edge system is indirectly affecting the OCPP 2.0 operations, due to its third-party nature. In the architectural schema, DSO represents this third-party organization. Similarly, an organisation or an individual that manages the EV charging system is refered to as Charging Point Operator (CPO). CPO is not depicted in the architecture, but it is assumed that the CPO's decisions are implemented by the CSMS.

*7) Energy Management System (EMS):* EMS is an intermediary system to the CSMS-CS communication. EMS controls a charging process by evaluating energy-based data fed by the consuming EV [15]. If an EMS is included in the EV charging system, the charging service may be supplied with electric power or with power provided by alternative sources, such as renewable energy sources. These alternative energy sources are also referred to as Distributed Energy Resources (DER). OCPP 2.0 supports the communication of the EMS with the CS and the reporting of the smart charging control limits that are imposed by the former to the latter [18].

*8) E-Mobility Service Provider (EMSP):* The role of the *EMSP* is to manage the economic arrangements regarding the EV charging service. EMSP issues the per EV or per EV driver contracts and manages the billing processes. It is common that the EMSP role is fulfilled by either the DSO or the CPO [25]. Either way, the DSO and the CSO inputs are processed by the CSMS; therefore EMSP is not depicted on the ARM schema.

| OSI model | | ISO 15118 | | |
|---|---|---|---|---|
| 7 | Application | OCPP, SDP, HTTP | | ISO 15118-2 |
| 6 | Presentation | SOAP/XML, JSON, XML/EXI | | |
| 5 | Session | V2G | | |
| 4 | Transport | TLS<br>TCP | UDP | |
| 3 | Network | IPv4 | IPv6 | |
| 2 | Data link | HomePlug Green PHY | | ISO 15118-3 |
| 1 | Physical | | | |

Fig. 2. OCPP in ISO 15118 communication stack [26],[27],[28],[29]

*9) Electric Vehicle (EV) user/driver:* Although not depicted in the ARM schema, the *EV user/driver* is an actor of the charging system [19]. Most implementations provide a user interface for the specification of the charging parameters or for the definition of the service reservations. The user via a smart device (e.g., smartphone) participates in the charging process configuration. The user's actions, the user device's vulnerabilities and the user's application add data and parameters to the service, hence indirectly affecting the OCPP operations and security. It is worth mentioning that several security methods for the identification of the consumer are based on the EV user's input.

### *C. OCPP communication and data sharing*

The development of OCPP started in 2009 and until today four versions have been released:

- OCPP version 1.2 [30] in 2011,
- OCPP version 1.5 [30] in 2012,
- OCPP version 1.6 [14] in 2015, and
- OCPP version 2.0.1 [13] in 2018

The latest version 2.0.1, also referred to as OCPP 2.0, which is the focus of this survey, has improvements providing the most complete set of 65 request/response messages for the CS-CSMS communication and improved security features. OCPP 2.0 supports the ISO 15118 standard for EVSE-to-EV communication [31]. ISO 15118 improves the previous standard IEC 61851 introducing the Plug & Charge (PnC) and the Smart Charging features [32]. OCPP supports these two features as well [33].

In the Open System Interconnection (OSI) and in the Transport Control Protocol/Internet Protocol (TCP/IP) models, the OCPP is an application-layer protocol. OCPP is depicted in the indicative list of the ISO 15118 compliant protocols in Figure 2. The Supply Equipment Communication Controller (SECC) Discovery Protocol (SDP) [27] is an alternative to the OCPP application-layer protocol. The protocols for the network, transport, session, presentation and application layers are described in the ISO 15118-3 part of the standard [34], whereas the protocols for the Physical and Data Link layers are described in the ISO 15118-2 part of the standard [31].

The OCPP exchanged messages are formatted based on the Simple Object Access Protocol (SOAP)/Extensible Markup Language (XML) or on the JavaScript Object Notation (JSON) standard, with the latter being the preferred choice for the presentation layer standard [25]. JSON is preferred because it provides messages that are smaller in size and faster to process [27].

The operations in the session layer are fulfilled by the Vehicle-to-Grid (V2G) protocol [35]. V2G manages the end-to-end sessions for the respective application-layer protocol. In the encapsulation process, V2G adds an 8-byte-long header to the payload, namely the Vehicle to Grid Transport Protocol (V2GTP) header. V2GTP is an identification factor for the V2G message when transferred in a byte stream [27].

The OCPP-based communication is supported by various transport layer protocols, such as Transport Layer Security (TLS), TCP, and User Datagram Protocol (UDP). TLS provides a secure communication through an encrypted channel for the transport layer end-to-end communication, and protects the application-layer payload [36].

### *D. Power trading management*

OCPP 2.0 complies with the ISO 15118 standard and for a complete power trading management it includes pricing and billing features. The ISO 15118 standard describes the PEV network charging service supporting the PnC feature, which allows the V2G communication to be held over a power line.

In an OCPP-based PEV network, the CS operates as a communication gateway between the EV and the back-end systems, such as the CSMS (Figure 3).

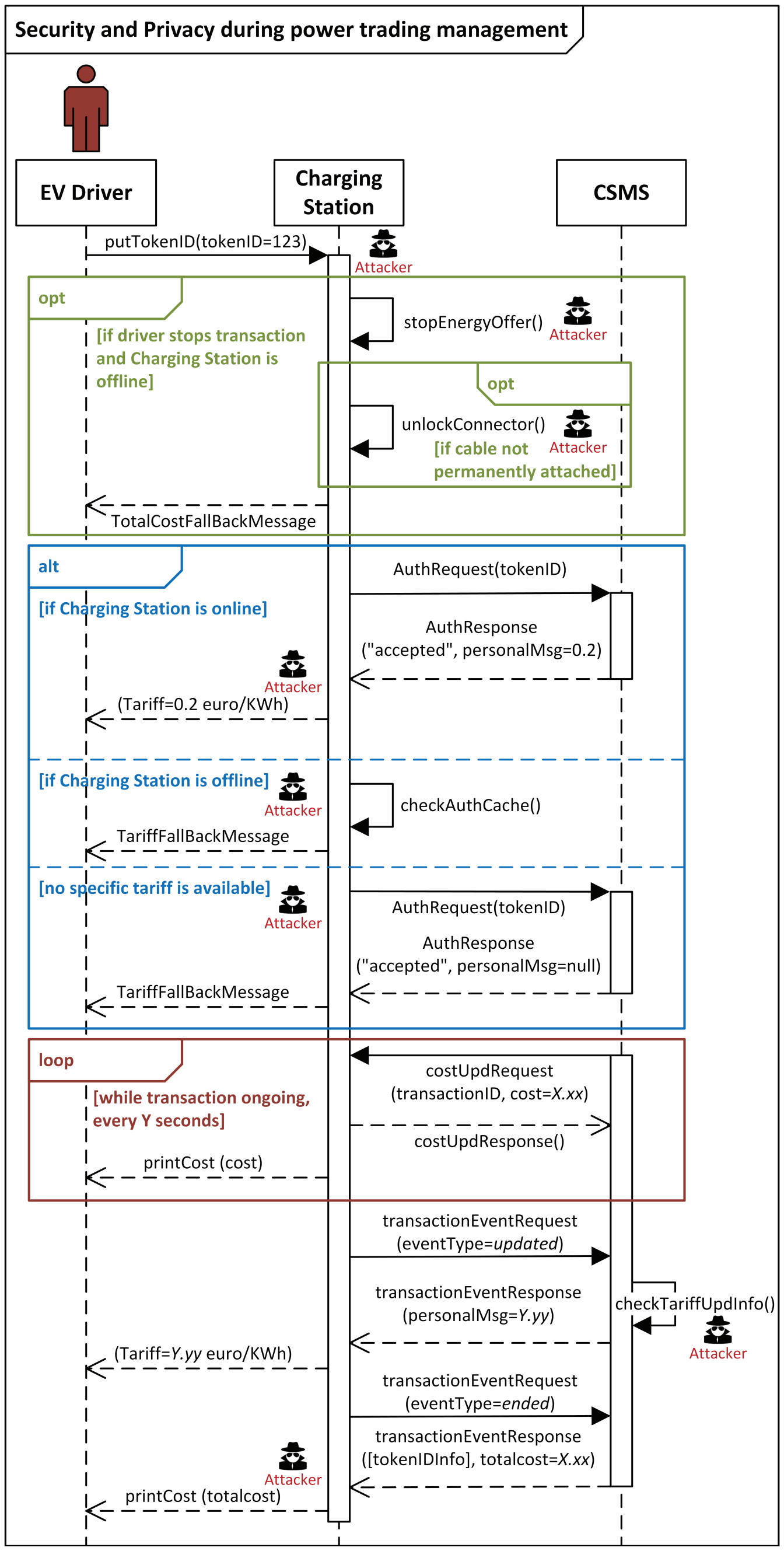

Fig. 3. Communication in power trading management process

Regarding the power trading management, CS collects the EV billing information [37]. The OCPP 2.0 power trading management may offer the EV driver the options of choosing the payment method, the preferred tariff or currency, and the charging mode [38].

OCPP 2.0 is the first protocol version that introduces the *TariffAndCost* functional block, which contains the messages and the datatypes for the power trade [15]. According to the OCPP specifications, this functional block considers a list of requirements aiming to protect *(a)* the billing process, *(b)* the privacy of the EV driver (e.g. driver tokenID), and *(c)* the communication between the CSMS and the CS. The *TariffAndCost* block allows the PEV network to provide the EV driver with information regarding the tariff cost of a charging service before the service starts, to update the tariff when changes apply, and to confirm the cost after the service is completed [33].

As shown in Figure 3, the billing information transferred within the PEV network raises certain security and privacy issues. The security of the exchanged data depends on the protocol's operations, and the level of billing data protection when stored, as well as when transmitted.

### *E. Co-existing protocols within a PEV network*

OCPP was designed to be interoperable with other protocols within the PEV network. In an OCPP-based PEV network, the end-to-end communication may be covered by other protocols, as well. The outcome of that communication, however, is a variable in the OCPP communications. Figure 4 shows the protocols and standards that co-exist with the OCPP within a PEV network. These protocols and standards are shown with a reference to the ISO 15118 protocol stack and are then described with a reference to the de jure OSI model.

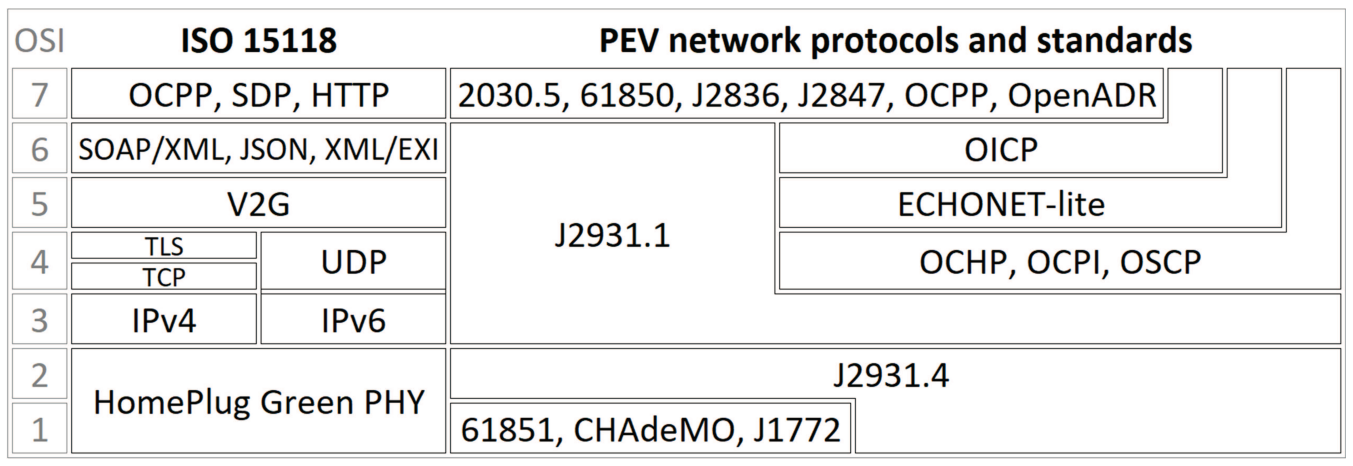

Fig. 4. Protocols and standards per OSI layer

*1) Physical layer:* The IEC 61851 standard describes an Alternative Current (AC) or Direct Current (DC) conductive charging system. It also describes the utilization of the control pilot signal [31] for the EVSE-EV communication. More specifically, the control pilot signal is used to verify the uninterruptible connection of the EV before the data transmission data as well as during the charging process. Regarding the CS-EMS communication, neither IEC 61851 nor IEC 61850 support the reverse power flow from the EV to the grid [3].

SAE J1772 describes the EVSE-CS fast charging communication [39]. J1772 or Type 1 is one of the common types of connectors used in the U.S., with the other two being the CHAdeMO, and the Tesla combo types [40]. The J1772 connector supports the AC discharging of the EV and the CHAdeMO the DC discharging. The OCPP *CPPWMController* block, which is used for the control pilot Pulse Width Modulation (PWM), adopts the SAE J1772 low voltage DC and PWM signaling [15].

As already mentioned, CHAdeMO is, similarly to J1772, a bidirectional flow standardised connector for the connection of the EV on the EVSE interface of a CS. CHAdeMO is the research product of Nissan and a Japanese hardware vendor, Nichicon. CHAdeMO supports the OCPP-based EVSE-EV communication [41]. The OCPP *CHAdeMOCtrlr* block manages the configuration related to the CHAdeMO controller [15].

The SAE J2931 standard includes the requirements of the CS-EV communication. The standard is the SAE equivalent to the IEC 15118.2 and to the IEC 15118.3 with the J2931.1 and the J2931.4 parts respectively. J2931.4 defines the broadband Power Line Communication (PLC) of the CS and the EV in L1 and L2, whereas the J2931.1 defines the protocol for the OSI L3 to L6 [26]. The standard covers the PLC alternative of the orthogonal frequency division multiplexing (OFDM)-based power line carrier [9].

*2) Data Link and Network layers:* It should be noted that the SAE J2931 is the one standard other than the ISO 15118, that refers to the OSI L2 and L3.

*3) Transport layer:* The Open Smart Charging Protocol (OSCP) is an OSI L4 to L7 protocol for the DSO-CPO communication [25]. OSCP was first introduced as an alternative to the OSI L7 OpenADR protocol, but it was soon expanded to cover all L4 to L7 layers. The same layers are covered by the CPO-EMSP roaming protocols, namely the OCPI and the OCHP.

The Open Charge Point Interface (OCPI) is an OSI L4 to L7 protocol allowing EVs to charge in various PEV networks. OCPI is JSON-based. TLS is optional for the OCPI protocol [25].

The Open Clearing House Protocol (OCHP) allows connections between the Clearing House (a role usually fulfilled by the CSMS, offering a CPO-EMSP communication platform in a cross-site manner), the EMSP and various CPOs. OCHP is an OSI L4 to L7 SOAP-based protocol, for which TLS is optional[25].

*4) Session layer:* The Japanese Echonet Consortium proposed the ECHONET-lite [42] protocol. ECHONET-lite is focused on the CS-EMS communication, but not exclusively [43]. Compared to OCPP, ECHONET-lite is an OSI L5 to L7 protocol [44] that utilizes binary-formatted messages. Data security is not in the protocol's consideration [45]. However, ECHONET-lite supports PLC, Bluetooth, Ethernet, and wireless Local Area Network (LAN) among others [46].

*5) Presentation layer:* The Open InterChange Protocol (OICP) [47] is an OSI L6 and L7 protocol that supports the same as OCHP communication, but it is both JSON- and SOAP-based. OICP is also referred to as a roaming protocol.

*6) Application layer:* In 2020, ISO and IEC introduced the ISO/IEC 15118 standard for the communication between the entities of an EV charging system [31],[32],[34],[48],[49],[50],[51]. ISO/IEC 15118 describes the V2G communication protocols per OSI layer. The standard introduces the PnC feature, which enables the authorized EV charging process. OCPP has been developed following this standard.

The SAE J2836 standard includes the use cases, and the SAE J2847 standard includes the specifications for the V2G communication in the OSI L7. J2836 and J2847 are the SAE equivalents to the IEC 15118.1 [26].

IEEE adopted the Smart Energy Profile (SEP2) protocol from the ZigBee Alliance, and proposed the evolved IEEE 2030.5 protocol [52]. IEEE 2030.5 is an OSI L7 IP-based protocol [45] that covers the EVSE-EV and the DSO-EMS communication, where DSO is the third-party operator [53]. The protocol's messages are XML or EXI formatted, and TLS is implemented in OSI L4 for data security. Regarding the certification process, IEEE 2030.5 supports life-long unrevoked private certificates as opposed to Public Key Infrastructures (PKI) [54]. So far, the protocol has been used only in US projects.

IEC 61850 is a standard that models the integration of the charging infrastructure with the smart grid [45]. IEC 61850 was developed by the CHAdeMO Association as an OSI L7 alternative to IEEE 2030.5 and to OCPP for the CS-EMS communication [55]. The standard proposes an intra-CS local network to reduce wiring [56]. The standard also supports the CSMS-DSO, the CS-CSMS, and the CS-CPO communications. OCPP and IEC 61850 complement each other for the EVSE-DER communication, as they consider the alternative energy sources that may exist within the EV charging system [57]. The DER communication systems are described in detail in the IEC 61850-7-420 [58].

OpenADR, formerly IEC 62746-10-1, is an OSI L7 [59] open demand-response protocol provided by the OpenADR Alliance. OpenADR is developed to support the CSMS-DSO communication [60]. In some cases, OpenADR has been used for the DSO-CPO [61] and for the DSO-EMS communications [53]. The compliance to this standard enables peak power demand reduction, load shifting, and dynamic pricing [1]. In terms of security, TLS is mandatory in the transport layer of OpenADR. Additionally, the OpenADR authentication is based on PKI and trusted certificate authorities (CA) [54]. OpenADR combined with the OCPP gives a dynamic nature to the CSMS with real-time adjustments to grid events.

## III. Related Surveys

The existing surveys regarding the security of the EV charging systems are presented in this Section. Table II depicts the *(a)* the architectural model every study provides and, hence, the set of assets under study, *(b)* the set of security countermeasures provided, if any, *(c)* the depiction of an association between the security threats and the affected assets, *(d)* the study of the OCPP security issues, and *(e)* the discussion topics for security open issues. Table II also indicates what was studied, partially studied, or not considered by the related surveys in the literature, as well as a comparison between these surveys to the current.

In an early 2016 paper, Han and Xiao [62] surveyed issues related to privacy preservation problems in V2G networks. They presented solutions that try to protect data privacy along with typical attacks against privacy. Moreover, they presented unsolved issues and possible countermeasures that were based on the existing solutions. The article provided a complete analysis on privacy preservation issues related to V2G applications. They mainly discussed issues regarding the location privacy, the ID privacy, the anonymous authentication,

TABLE II
SURVEYS ON EV CHARGING SYSTEMS AND OCPP SECURITY

| Ref | Year | Architectural model | Countermeasures | Threats-assets association | OCPP versions | Security open issues |
|---|---|---|---|---|---|---|
| [62] | 2016 | V2G network | ✓ | - | - | Data security |
| [63] | 2017 | EV charging infrastructure | - | - | - | V2G network security |
| [64] | 2017 | OCPP-based charging point datagram | ✓ | (✓) | 1.6 | OCPP v1.6 security |
| [25] | 2020 | Dutch EV-charging infrastructure | - | - | 1.6, 2.0* | EV driver's authentication |
| [39] | 2020 | EV charging system | (✓) | (✓) | 1.6, 2.0* | (a) OCPP security concerns;<br>(b) symmetric or asymmetric cryptography;<br>(c) enforced authenticity; and<br>(d) infrastructure security |
| [54] | 2021 | EV ecosystem front- & back-end protocols | - | (✓) | 1.6, 2.0* | EV-charging dedicated PKI |
| [65] | 2021 | EV CSs into electricity grids | ✓ | - | 1.6 | - |
| [66] | 2021 | EV charging station architecture | (✓) | - | 1.5, 1.6 | - |
| This | 2021 | OCPP-based EV charging system | ✓ | ✓ | 1.2, 1.5, 1.6, 2.0* | ✓ |

[✓]: Studied; [(✓)]: Partially studied; [-]: Not included.
*OCPP version 2.0.1 (aka 2.0)

and the billing privacy. Even though this survey considered the security and privacy issues of an EV charging system, the charging protocol vulnerabilities were not analysed. Their analysis was more focused on the V2G network architecture rather than the EV charging elements. Moreover, an association between the security and privacy threats and the assets of an EV charging system was not provided.

Bernardini, Asghar and Crispo [63] conducted a thorough research in identifying the security and privacy issues in vehicular communications. They classified their analysis into three categories: intra-vehicle, gateways, and inter-vehicle communications. Although they identified most of the entities that participate in an EV charging scenario, their analysis only included the security issues associated with the DSO, the CS, and the EV. Moreover, they did not provide an association between the security and privacy threats and the assets. The security issues of the OCPP protocol are only partially discussed.

The security issues that are associated with the OCPP version 1.6 were presented by Alcaraz, Lopez and Wolthusen [64]. Several vulnerabilities related to boot notification and to the use of the TLS protocol were analysed. Moreover, the paper tried to identify the impact of the identified threats on several assets and communicating systems. The challenges presented in [64] were related to the earlier OCPP version 1.6, which lacked some of the security features of the current version studied in this survey. Moreover, [64] focused on the OCPP-based communication between the EV and the CS, leaving aside assets such as the EV driver, the DSO, the LC, the LP, and the energy sensors/controls. However, an association between the security threats and the assets under study was presented and open issues referring to the then released version of the OCPP were highlighted.

In a recent article, Audel and Poll [25] provided an overview of the main roles and protocols in the use of EV charging in the Netherlands. They argued that the TLS protocol cannot cover the security and privacy requirements that the system has. They proposed that it would be feasible to add end-to-end, long-term authenticity and end-to-end confidentiality to the data exchanged among different entities of the system. The article is very interesting and has spotted security issues by studying the Dutch EV-charging infrastructure. However, the energy sensors/controls and the related security issues were not considered. Additionally, an association between the security and privacy threats and the assets was not provided, and the security issues of the OCPP protocol were partially discussed. The open issue that was identified is the weak EV driver's authentication.

Antoun et al. [39] presented a detailed report regarding EV charging systems focusing on security and privacy issues. They presented in a graphical way the main vulnerabilities that exist both in residential as well as in public charging infrastructures, and identified some security gaps which need to be addressed in the near future. However, the assets that were studied relatively to the security threats were the EV, the CS, and the communication medium, resulting in a limited threat-assets association list. Similar limitations affected the countermeasures enumeration, including only solutions protecting the communication between the CS and the CSMS, as well as between the CS and the EV. Open issues regarding the protocol's security features, the limited use of cryptography, the lack of enforced authenticity, and the security of the power grid infrastructure were discussed.

Metere et al. [54] reviewed security and privacy issues related to the EV charging ecosystem focusing on smart charging and V2G applications. Several recommendations and guidance for securing the EV charging infrastructure were first presented, followed by a detailed analysis of the security and privacy issues that such a complex system faces. It was argued that security technology is already there but it needs to be adapted in order to take into account the specific challenges of the EV charging infrastructure. According to [54], a balance between QoS and security needs to be achieved. Issues that are associated with the OCPP were only briefly presented. Moreover, some assets and their security issues, such as the EV driver, were not studied while the CSs and the association between threats and assets were partially studied. The paper analysed in-depth the implementation of a Public Key Infrastructure (PKI) for EVs as a countermeasure for the EV ecosystem security issues. Metere et al. [54] referred to the need for an EV-charging dedicated PKI as an open future issue.

Pourmirza and Walker [65] reviewed the vulnerabilities and the cyber security issues of the EV CSs in the then current UK landscape. They emphasized on the disinformation attacks against the CSs that may expose the EV driver credentials and data. Beyond the CS, the security issues of three more assets were considered, namely the EV, the EV driver, and the CSMS. The security and privacy issues, as well as the countermeasures were studied for the OCPP version 1.6 supported CSs. A threat-asset association or a discussion about security open issues were not provided.

In the latest OCPP-related study, Raboaca et al. [66] focused on the application design for an OCPP version 1.6 supported EV charging system. They overviewed the related topologies and architectures and they analysed the OCPP operational features to develop a CS reservation application for the EV drivers. Although they overviewed the security issues on core assets, the security attacks were not associated with the assets that are mostly affected. They argued that the best practice for OCPP security enhancement is the integration of Blockchain and Artificial Intelligence (AI) techniques. For the future, the open issues discussed were on the topic of the CS performance stability and not on the security area.

## IV. OCPP Security Issues

The security and privacy issues of an OCPP-based EV charging system are discussed in this Section. The security requirements of the system, as well as of the elements operating in an EV charging system, are presented. Then, a taxonomy of the cyber attacks, the physical attacks and the cyber-physical attacks that may affect an EV charging system is introduced, based on the type and on the impact of every attack. For every attack, the proposed in the literature countermeasures are enlisted, and the effect of the attack and of every countermeasure per asset is depicted.

### A. Security requirements

The security requirements for the OCPP-based EV charging service include the integrity, the authenticity, the confidentiality, and the availability in the context of the EV driver's information, the EV's data regarding the State of Charge (SOC), the power micro-grid, and the billing process of the service.

*1) EV driver's information:* Regarding the billing process and the EV driver's information, non-repudiation and accountability are also important to ensure that every service cycle will be paid for, and all billing-related messages/notifications are sent and received by the proper recipient [67].

*2) EVSE:* The growth of the PEV use motivated ElaadNL to study the security requirements for a PEV network and all the participating devices and elements. According to the outcome, the security requirements for the EVSE are [16]:

*(a)* remote updates/upgrades option;
*(b)* limited down-time cases and periods; and
*(c)* cross-vendor interoperability.

The contractual design of the EVSE must include the option of remotely updating or upgrading the software or the firmware of the EVSE or of the CS. The lack of this option in combination with the fact that CS are scattered in public areas and, usually, away from the immediate access of an IT operator, could lead in delayed updates and, hence, in security issues.

The configuration of an EVSE must be easily changed without the need for a reboot. The same requirement exists for any firmware updates, which, may also, call for a reboot. This option would allow the frequent adjustment of the configuration without the communication loss during the reboot process. For the CS, which is always accessible by customers and which should have a minimum down-time and an updated configuration, this is a crucial feature operational- and security-wise. The communication between the CS and the back-end systems should be as uninterruptible as possible.

The remote unified migration of configuration parameters should be a cross-vendor capability, so as to ensure the interoperability of the various CS models within one PEV network. For the same reason, no information should be hardcoded on the CS by the vendor.

*3) Back-office systems:* The list of requirements for a back-office system includes the following:

*(a)* managing a network of multi-vendor devices;
*(b)* supporting smart charging; and
*(c)* resolving operational issues of multi-vendor devices.

*4) OCPP:* The OCPP documentation quotes the requirements considered for the OCPP 2.0 security block construction [15]. This list, which is a refined and targeted selection from the previously mentioned lists and the ElaadNL study regarding the security requirements on EV charging systems [68], consists of the following:

*(a)* The secure connection between the CSMS and the CS, including the cryptographic methods to ensure the integrity and the confidentiality of the messages.
*(b)* The mutual authentication between the CSMS and the CS.
*(c)* The secure firmware update process for the CS with ensured firmware images in terms of integrity and non-repudiation.
*(d)* The logging/monitoring of the smart charging service.

### B. Attack classification

The categorization of the OCPP-related attacks is based on the type and on the impact of the attack. Both the type and the impact of an attack on a Cyber-Physical System (CPS), such as an EV charging system, can be physical, cyber or both [69]. The *type of an attack* is physical, cyber, or cyber-physical depending on the way of deployment, whilst the *impact of an attack* is physical, cyber, or cyber-physical depending on the consequences that the deployment carries. For instance, a cyber-physical type of attack is the one deployed with both physical access and the appropriate cyber tool, and an attack with a cyber-physical impact is the one affecting both the cyber realm and the physical infrastructure.

Based on these criteria, the attacks and the relevant countermeasures already developed and tested on OCPP-based PEV networks can be categorized in three groups: *(a)* the physical attacks, *(b)* the cyber attacks, and *(c)* the interdependent cyber-physical attacks. In the first group, we classify the attacks that need physical access to a PEV network site or any architectural element to be unleashed. In the second group, we classify the attacks on PEV networks that are unleashed in a cyber manner without any physical access needed, and in the latter group the attacks that are deployed in a cyber-physical manner. This taxonomy is shown in Table III. The description, countermeasures, and affected EV charging assets for the physical attacks, the cyber attacks, and the cyber-physical attacks follow in the next Sections.

TABLE III
TYPE-BASED TAXONOMY OF OCPP-RELATED ATTACKS

| Type | Attacks | Section.Idx |
|---|---|---|
| **Physical** | Tampering | IV.C.1 |
| | *EVSE/CS tampering* | IV.C.1a |
| | *EV tampering* | IV.C.1b |
| | *Reservation/billing data tampering* | IV.C.1c |
| | *Multiple assets tampering* | IV.C.1d |
| | Side-channel | IV.C.2 |
| | State/sensor | IV.C.3 |
| **Cyber** | Man-in-the-middle (MitM) | IV.D.1 |
| | Packet replay | IV.D.2 |
| | Denial-of-Service DoS/DDoS | IV.D.3 |
| | ARP spoofing | IV.D.4 |
| | Remore Keyless Entry (RKE) cloning | IV.D.5 |
| | Malware | IV.D.6 |
| **Cyber-physical** | Power outage/overload | IV.E.1 |
| | *Protection of the micro grid* | IV.E.1a |
| | *Protection of the power grid* | IV.E.1b |
| | Substitution | IV.E.2 |
| | Meter bypassing | IV.E.3 |
| | Over-the-Air (OTA) tampering | IV.E.4 |
| | Smart card cloning | IV.E.5 |
| | Masquerading & impersonation | IV.E.6 |
| | False Data Injection Attacks (FDIAs) | IV.E.7 |
| | Insider attacks | IV.E.8 |
| | Switching attacks | IV.E.9 |

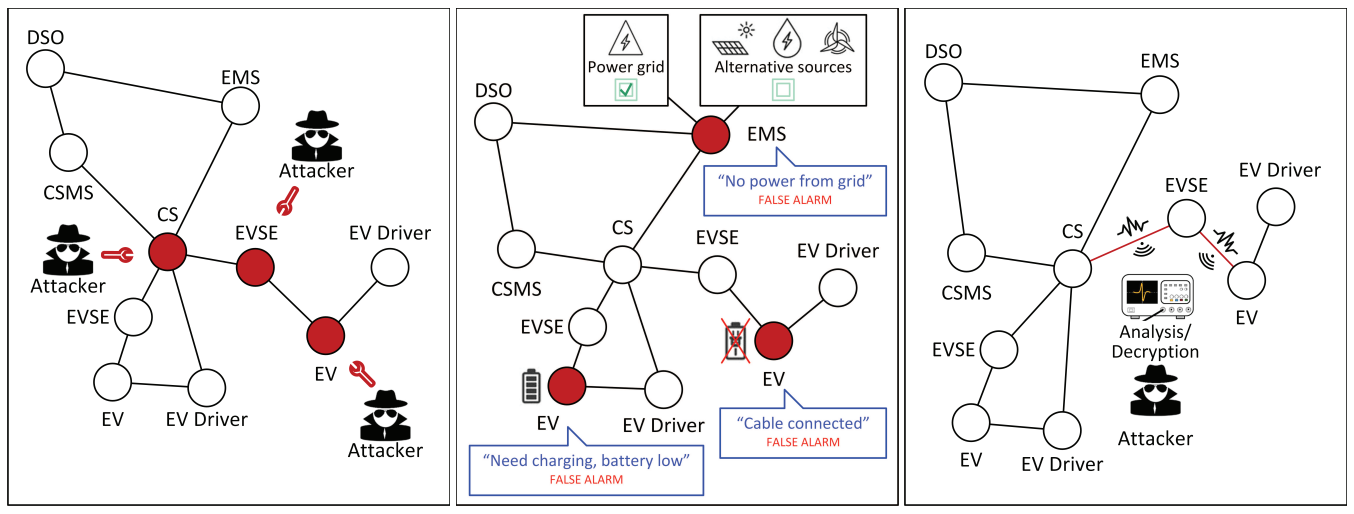

(a) Tampering (b) Side-channel (c) State/sensor

Fig. 5. Physical attacks on PEV network

## C. *Physical attacks*

The first group of attacks, i.e. the physical attacks (Figure 5), and the related countermeasures that are proposed in the literature are described in this Section. This group contains the attacks that are either physically unleashed or that have physical consequences on the assets of an EV charging system.

In Table IV, all the attack-related countermeasures are shown, followed by the assets of an EV charging system. If an asset is targeted by the attack and the proposed countermeasure is protecting the asset against this specific attack, then the asset is marked as *Protected asset* in the Table. Otherwise, the asset is marked as *Targeted asset*. If the specific attack does not affect the asset, then this asset latter is not marked in any way.

*1) **Tampering**:* Physical access control can protect not only the scattered elements of the EV charging system, such as the EVs, the driver's smart device, and the EVSE, but also the rest of the system's assets. These elements are particularly exposed to physical access by being installed or parked in publicly accessed areas. In addition to this exposure, each of these elements has operational or constructional characteristics that may facilitate physical access attacks. The following countermeasures are grouped based on the EV charging element that is used as a tampering entry point.

*a) EVSE/CS tampering:* The EVSE is the edge device hosted in the edge element namely the CS. EVSE is accessible to the EV driver and in some cases to anyone, when the CS is located in public charging sites. Additionally, most EVSEs have external ports and USB or serial interfaces for the EV connection; ports that allow and may be exploited by a physical access attack, thus affecting the EVSE operations, software, and hardware or the personal data of the EV driver. EVSE is vulnerable mostly because of the included cellular modem, which is used for the exchange of the card reader's data between the EVSE and the card issuer [91]. In case the modem is tampered, then the payment process may be unreliable, the card owner credentials may be exposed, and the charging service may never be fully completed.

As proposed in the report authored by the National Renewable Energy Laboratory for the U.S. Department of Energy (DOE) [36], the constructional design of the EVSE can mitigate the physical access attack types by *(a)* reducing the number of external interfaces, *(b)* enforcing encryption methods for the incoming data, and *(c)* embedding tampering alarms and auditing techniques. OCPP 2.0 incorporates cryptographic signature and firmware update features [71]. For that reason, OCPP is suggested as the optimal protocol against the CS tampering and, more specifically, the EVSE tampering [39].

EVSE is not the only CS asset that may be tampered. The CS holds public and private certificates for the authenticated communication with the EV and the CSMS. An anti-tampering solution is the use of a smart card chip for the generation and for the storage of the digital signatures [72]. If the smart card chip gets tampered with, then the CS will be alerted, and the relative certificate will be revoked by the CSMS. In addition to these features, modern smart cards embed micro-processors allowing them to also host cryptographic operations [73]. The use of a smart card as an anti-tampering solution has been, also, proposed in [74] for the protection of the EV driver's data, during the payment process.

*b) EV tampering:* As the endpoint of the charging service, every EV is identified and authenticated by the service's CSMS. The EV's identification and authentication is, usually, based on validation of the EV and of the EV driver's credentials. An EV tampering attack may result on a biased

TABLE IV
PHYSICAL ATTACKS ON OCPP-BASED EV CHARGING SYSTEM

| Attacks | Countermeasures | Assets | | | | | | | |
|---|---|---|---|---|---|---|---|---|---|
| | | Driver | EV | EVSE | CS | EMS | CSMS | Data | Grid |
| Tampering | Constructional EVSE design [36] | ○ | ○ | ● | ○ | ○ | ○ | ● | ● |
| | In-vehicle credentials generation/storage [37],[70] | ○ | ● | ○ | ○ | ○ | ○ | ● | |
| | OCPP encryption and firmware updating [71] | ○ | ○ | ● | ● | ○ | ○ | ○ | |
| | Smart card chip for signatures [72],[73],[74] | ○ | ○ | ○ | ● | ○ | ● | ● | |
| | Intelligent electronics device on EV [75] | ○ | ● | ○ | ○ | ○ | ○ | ○ | |
| | Limited lifetime of EV authentication [76] | ○ | ● | ○ | ○ | ○ | ○ | ○ | |
| | OCPP PnC mechanism [1] | ○ | ● | ○ | ○ | ○ | ○ | ○ | |
| | Elliptic-curve keypair for AutoCharge [77] | ● | ● | ● | ○ | ○ | ○ | ● | |
| | Decentralized firmware attestation [78],[79] | ○ | ● | ○ | ○ | ○ | ○ | ● | |
| | Direct anonymous attestation protocol [80] | ○ | ● | ○ | ○ | ○ | ○ | ● | |
| | EV reservation w/ smart contract [81] | ● | ● | ○ | ● | ○ | ○ | ● | |
| | Authentication scheme w/ smart contracts [82] | ● | ● | ○ | ● | ○ | ○ | ● | |
| | Pseudonym-based authentication scheme [83],[84] | ● | ● | ○ | ● | ○ | ○ | ● | |
| | Physical unloneable functions [85],[86] | ● | ● | ● | ○ | ○ | ○ | ○ | |
| | Secure user key-exchange authentication protocol [87] | ● | ● | ● | ○ | ○ | ● | ○ | |
| Side-channel | In-vehicle credentials generation/storage [37],[70] | ○ | ● | ○ | ○ | | | ● | |
| | Dedicated in-vehicle hardware [88],[89] | ○ | ● | ○ | ○ | | | ● | |
| | Physical security policies [55] | ○ | ○ | ○ | ● | | ● | ● | ● |
| | Decentralized smart charging controller [90] | ○ | ○ | ○ | ● | | ● | ● | |
| State/sensor | N/A | | ○ | ○ | ○ | ○ | ○ | ○ | ○ |

○ Targeted asset ● Protected asset

set of credentials or on the exposure of the EV, of the CS or of the EV driver's sensitive data. A tampered EV may distort the payment/billing process or mislead the CS or the CSMS with false charging metrics and indications.

To address the EV authentication Chan and Zhou [75] assumed that every EV should be equipped with an onboard Intelligent Electronics Device (IED) to serve as an authentication token for the EV within the EV charging system. The IED must be tamper-resistant and the key that is stored in it should be accessed, revoked and issued only by the operator of the charging system (i.e. the person or the authority operating the CSMS). The necessary adjustments for the EV to include the IED is considered easy and of low cost.

Based on the IEC 61850 standard and on the expected adoption of Software Defined Networks (SDNs), Soares et al. [76] introduced an authentication mechanism for EVs with limited control load and authentication lifetime.

The latest OCPP version 2.0 supports the PnC mechanism for the identification and authentication of the EV [1] via the charge cable. PnC is aligned with the ISO/IEC 15118 standard and it allows the EV's authentication without human intervention. However, the PnC mechanism carries a vulnerability caused by the transmission of the EV's credentials back and forth between the CS and the CSMS during the authentication process. A mitigation measure for this issue is the generation and the storage of the credentials within the EV [37],[70].

An alternative to the PnC is the Autocharge mechanism [92]. Autocharge is supported by the previous to the OCPP 2.0 versions, however several existing implementations are supported by those OCPP versions, hence related security issues still need to be addressed. Autocharge includes the vehicle-to-charger ISO/IEC 15118 Signal-Level Attenuation Characterisation (SLAC) protocol. SLAC is a demand/response protocol used when the EV and the CS share a PLC connection. The EV and the CS agree on a unique ID key per session and any data exchange amongst them embeds this key. The SLAC ID is provided by the CS and by the PKI service. In addition to the case that the PKI service is inaccessible, the SLAC initialization messages are exchanged in plain text, leaving the ID and, consequently, the session exposed to eavesdropping attacks. Baker and Martinovic [77] propose the use of a temporary Elliptic-Curve keypair for the SLAC initialization messages, until the session's ID key is produced and becomes known to both parties.

Another challenge in the case of EV tampering is the integrity of the Electric Control Unit (ECU) of the EV. Although an ECU tampering may have the goal of altering the charging process, ECU controls far more critical tasks that may also be affected, such as the engine control. ECU controls the charging process of the EV and, therefore, a man-in-the-middle (MitM) attack on the OCPP may ultimately affect the ECU. In [78] and [79] a decentralized firmware attestation scheme is proposed to detect the tampering of the ECU flash-memory or a stale firmware on the ECU.

The ISO 15118 communication between the EV and the CS is accomplished from the EV side via the Electric Vehicle Communication Controller (EVCC) embedded device. EVCC handles the communication and, therefore, it carries sensitive data regarding the Original Equipment Manufacturer (OEM) Provisioning and Contract Certificates and the related keys of the EV. Zelle et al. [80] propose the use of the Direct Anonymous Attestation (DAA) protocol and of the in-vehicle Trusted Platform Module (TPM) to protect the EVCC and the data against tampering throughout the PnC process chain. TPM protects the keys and certificates by hosting their generation and storage processes, and DAA supports the encryption and authorization processes during the charging service's life cycle.

*c) Reservation/billing data tampering:* The tampering of the EVSE/CS, the EV, or of any other element of the EV charging system aims to, ultimately, intercept or tamper with the data and, therefore, to control the reservation or the payment/billing process. The data tampering is strongly associated with energy theft, not only of the charging system's micro-grid but of the local grid, as well [64].

The BlockEV [81], a blockchain-based EV charging protocol, supports the EV-CS communication without any private information sharing from either side. This is accomplished with the use of smart contracts and a distributed ledger for the relevant to the contracts data [93],[94]. The BlockEV features ensure:

*(a)* the availability of the CS chosen for reservation;
*(b)* the validity of the reservation placed by the EV;
*(c)* the credibility of the EV or the EV driver;
*(d)* the price for the reserved service;
*(e)* the identity of the EV actually served; and
*(f)* the amount of consumed energy.

The smart contracts are used for the privacy-preserving authentication scheme in [82]. In this case, the authentication process uses the Pederson Commitment and the token-based mechanism to be completed anonymously.

The EV and the EV driver's data protection against tampering is studied in [83]. This work proposes an authentication scheme for the communication between the EV and the CS that provides privacy regarding the EV's and the driver's data. For that reason, the identification of the EV is based on a pseudonym produced for every EV-CS pair in the EV charging system. The pseudonym changes if the EV connects to a different CS and expires when the EV leaves the charging area. Only a central element, such as the CSMS, needs to keep record of any change or any expiration of the existing pseudonyms. The Portunes protocol [84] proposes the authentication of the EVs with a pseudonyms mechanism, that provides location privacy for the EV as well. Portunes was compared against the Elliptic Curve Digital Signature Algorithm (ECDSA) signature generation and verification and it was found to be much faster.

*d) Multiple assets tampering:* Any tampering of the EV, of the driver's smart device, or of the EVSE can be mitigated by embedding the Physical Unloneable Functions (PUFs), as proposed by [85],[86]. PUFs alter the device's behavior, making the tampered device identifiable within the EV charging system. Once the tampered device is identified, it can be more easily contained. The secure user key-exchange authentication protocol named SUKA [87] is also based on PUFs. SUKA achieves a two-step mutual authentication between an EV and the grid server. SUKA can provide session key security, physical security, message integrity, and identity protection. In addition, SUKA can protect the V2G communication against impersonation, replay, and MitM attacks.

*2)* ***Side-channel attacks****:* The participation of the OCPP communicating elements in the charging and billing processes introduces a risk associated with side-channel attacks for those elements and, hence, for the protocol itself. The power analysis attack is a hardware attack, and one of the main classes of the side-channel attacks. This attack may affect the EV charging system's elements that produce or host sensitive information and, more specifically, the secret keys and the credentials [95]. The most vulnerable to power analysis elements are *(a)* the EV driver, *(b)* the EV, *(c)* the EVSE, and *(d)* the CS. All these key charging elements not only produce, store or exchange messages containing sensitive information but, also, they include hardware that can be monitored regarding its power consumption level. Depending on the targeted element and on the acquired credentials and keys, the power analysis may lead to impersonation, billing fraud, or meter bypassing attacks [80].

OCPP is also affected by the electromagnetic attack, a side-channel class based on leaked electromagnetic radiation, in the cases where a PLC exists between the EV and the CS [77]. The PLC circuit is by design operating as an antenna and the waveforms of a PLC communication can be wirelessly eavesdropped and manipulated with ease.

The side-channel attacks mostly focus on obtaining either the EV's or the driver's credentials. These credentials are not only useful to the attacker, but are, also, exposed to attacks. The OCPP PnC mechanism is vulnerable to side-channel attacks, because of the transmission of the EV credentials between the CS and the CSMS, and vice versa, during the authentication process [80].

Most side-channel countermeasures aim at protecting the affected assets rather than preventing an attack against them. A mitigation measure for this issue is the generation and storage of the credentials within the EV, as proposed in [37],[70]. The generation and storage are carried out by a Hardware Security Module (HSM) embedded for this purpose in the EV. The relevant messages exchanged between the CS and the CSMS are natively supported by the OCPP 2.0 in the form of *DataTransfer* messages. This dedicated hardware chip implementation is preferable to the System on Chip (SoC) alternative, in terms of operational independence and ease of access.

An earlier dedicated hardware chip implementation, TPM 2.0 [88], has already been integrated in EVs that are in use. TPM 2.0 supports a secure data and certificate keys storage, cryptographic and authorization operations, and a detection method for firmware manipulation. TPMs are aligned with the AEC-Q100 standard and they conform to the Common Criteria (CC) EAL4+ security certification. Fuchs et al. [89] propose a secure architecture that exploits the security operations of the TPMs.

Even in the case of identification and authentication of both the EV and the driver, sensitive information can be processed on-site instead of being transmitted to the CSMS and back. This schema may address the secure communication requirements of the PEV network. However, the extra devices per CS needed for distributing the authentication process introduce their own additional physical security issues.

The implementation of physical security policies on these integrated devices is proposed to protect the integrity of the devices and the reliability of the power grid [55].

The project team working on an open-architecture software platform for EV smart charging proposes the implementation of an OCPP supported decentralized controller in every CS (i.e. in the XBOS-V controller) [90]. The scattered decentralized controllers are coordinated by the CSMS and relieve both the CSMS and the CSs from the computational overload. The CSs do not need to carry networking hardware interfaces or perform complex communication processes.

*3) **State/sensor attacks**:* As PEV network is highly dependable on sensors, a service can be affected by malicious sensor values in many ways. The sensor attack is usually performed to ultimately lead to a False Data Injection Attack (FDIA) [96], a jamming of the sensor's communication with the ECU, a GPS deception or even a Denial of Service (DoS) of the ECU [97]. All these consequent attacks may also affect the intra-ECU or the extra-EV communications and, therefore, the OCPP. To the best of the authors' knowledge, the literature does not contain a work or study regarding a countermeasure specifically for this physical attack.

### D. *Cyber attacks*

The second group of attacks, the cyber attacks (Figure 6), and the related countermeasures that are proposed in the literature are described in this Section. This group contains the attacks that have cyber consequences on the assets of an EV charging system.

In Table V, the countermeasures related to every attack and the assets of an EV charging system are shown. The assets are either not marked if the attack does not affect them, or marked as *Targeted asset* if the attack affects them and no countermeasure exists, or marked as *Protected asset* if the countermeasure is protecting the asset.

*1) **Man-in-the-Middle (MitM) attack**:* The MitM in the PEV network context can be deployed by having as entry points characteristics or vulnerabilities of one or more elements. One of these cases is the exploitation of the CS USB ports, which are usually publicly accessed. This access to the CS USB ports allows the attacker to obtain logs and critical data of the CS itself, and of the EV. Also, these attacks may result in installing malicious firmware and software on the CS or in changing the CS clock and in distorting the service [39]. In most of the cases, however, MitM attacks are targeting the CS-CSMS or the EV-CS/EVSE communications [36] as well as the data exchanged.

The EVs are also a preferable entry point for MitM attacks. These attacks exploit ECU vulnerabilities to take control of the EV and then distribute the attack to the PEV network. The decentralized firmware attestation scheme proposed in [78],[79] uses the ECU flash-memory and firmware protection to mitigate those vulnerabilities.

The MitM attacks on the communication between the CS and the CSMS may leak the information exchanged regarding the prices, the firmware, and the access control policies among others. The protection of the CS-CSMS communication against MitM is studied in [100], where the OCPP *DataTransfer* operational message is used to exchange important data, such as the meter value, in an encrypted manner. Additionally, the data is segmented in shares and each share is transmitted in separate messages. Thus, an MitM attacker would not only need to decrypt but, also, to collect enough shares to be able to recover the original data.

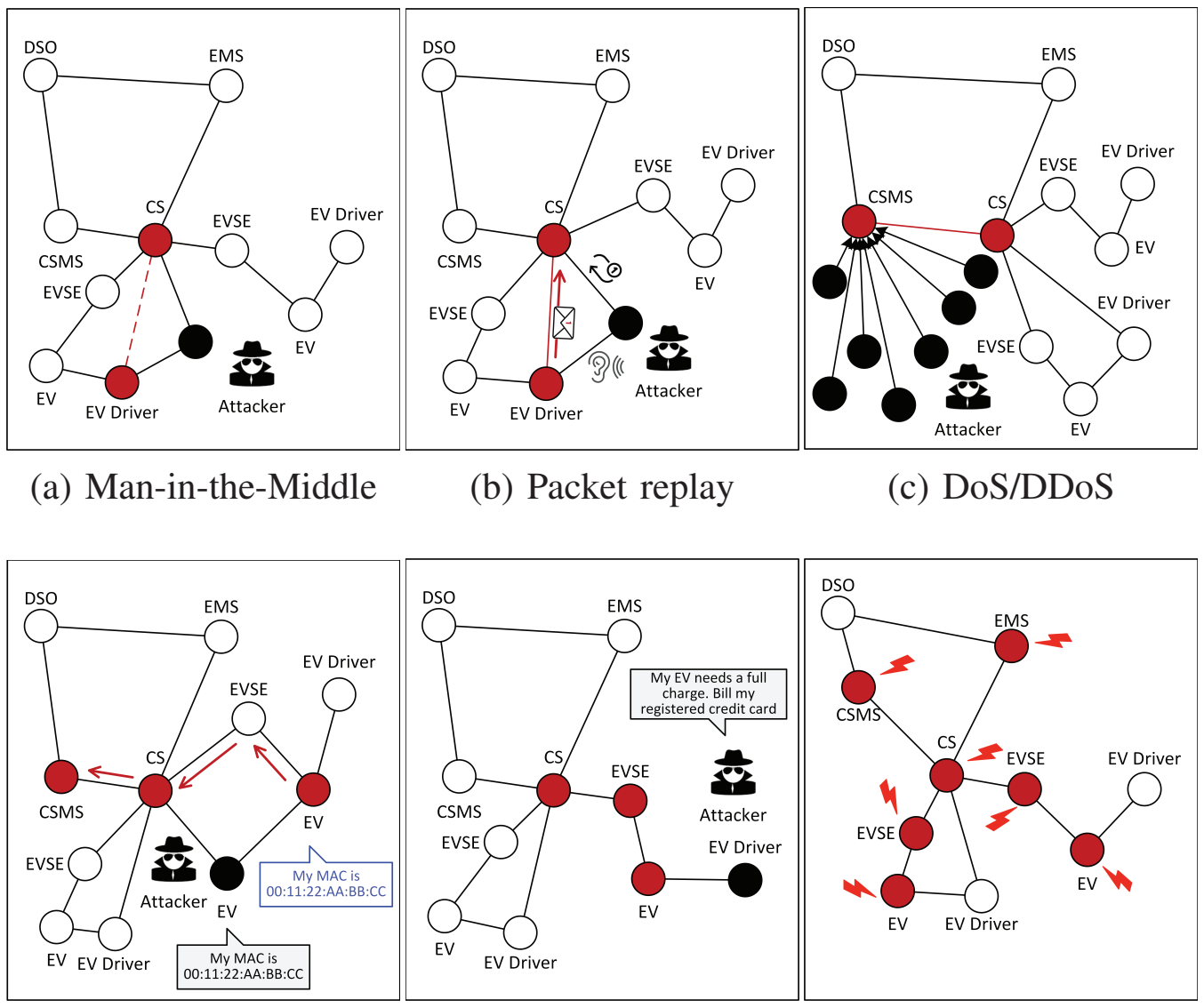

(a) Man-in-the-Middle (b) Packet replay (c) DoS/DDoS

(d) ARP spoofing (e) RKE cloning (f) Malware

Fig. 6. Cyber attacks on PEV network

The MitM on the EV-EVSE communication may leak the location of the EV, financial information of the service or the EV driver, and security/identity credentials of the EV or the driver. If such an attack succeeds and the attacker manages to set up connectivity, for example with the EVSE, the attacker may proceed and make the EVSE redirect traffic to itself (i.e. sinkhole), or to randomly select hosts and deploy a distributed attack (i.e. wormhole) [64].

To address this issue, the Multimodal and Multi-pass Authentication using Contract Certificate (MMA-CC) scheme [101] was proposed, where a charging service is activated only for an EV with a valid contract certificate submitted alongside the EV driver's credentials. The same communication can be protected by the EV power trading decentralized architecture based on the consortium blockchain [102], [103], where the trader's digital signature is embedded on the data and then encrypted with the recipient's public key. The Direct Anonymous Attestation (DAA) protocol and the in-vehicle TPM [80] are proposed to mitigate MitM for the communication between the Electric Vehicle Communication Controller (EVCC) and the Supply Equipment Communication Controller (SECC), by introducing encryption and authorization processes. Additionally, a blockchain-based EV charging solution can enable the EV-CS communication without sharing their private information but instead using a smart contract [81].

Regarding the EV-CS communication and the SLAC initialization messages these two ends exchange, the proposal

TABLE V
CYBER ATTACKS ON OCPP-BASED EV CHARGING SYSTEM

| Attacks | Countermeasures | Assets | | | | | | | |
|---|---|---|---|---|---|---|---|---|---|
| | | Driver | EV | EVSE | CS | EMS | CSMS | Data | Grid |
| MitM | Elliptic-curve keypair for Autocharge [77] | ● | ● | ● | ○ | ○ | ○ | ● | |
| | Decentralized firmware attestation scheme [78],[79] | ○ | ● | ○ | ○ | ○ | ○ | ● | |
| | Direct anonymous attestation protocol [80] | ○ | ● | ● | ○ | ○ | ○ | ● | |
| | EV reservation w/ smart contract [81] | ● | ● | ● | ● | ○ | ○ | ● | |
| | Three-factor authentication protocol [98] | ● | ○ | ○ | ● | ○ | ○ | ● | |
| | Back propagation neural network scheme [99] | ○ | ● | ○ | ● | ○ | ● | ○ | |
| | OCPP *DataTransfer* block [100] | ○ | ○ | ○ | ● | ○ | ● | ● | |
| | Multimodal multi-pass authentication w/ contract certificate [101] | ● | ● | ● | ○ | ○ | ○ | ● | |
| | Power trading decentralized architecture [102],[103] | ● | ● | ● | ○ | ○ | ○ | ● | |
| Packet replay | Message authentication code [71] | | ● | ○ | ○ | | | ● | |
| | Three-factor authentication protocol [98] | | ○ | ● | ● | | | ● | |
| | Abnormal behavior detection system [104],[105] | | ● | ○ | ○ | | | ● | |
| | BC-enabled EV charging system [106] | | ● | ○ | ● | | | ● | |
| | Distance bounding algorithm [107] | | ● | ● | ○ | | | ● | |
| DoS/DDoS | Three-factor authentication protocol [98] | ● | ○ | ○ | ● | ○ | ○ | ● | |
| | Power trading decentralized architecture [108] | | ● | ● | ● | ○ | ● | ● | ● |
| | Traffic detection, back propagation neural network [109] | ● | ○ | ○ | ● | ○ | ○ | ○ | |
| | BC-based EV bidding protocol [110] | ● | ● | ○ | ● | ○ | ○ | ● | |
| | BC-based distributed ledger technology [111],[112] | ● | ● | ○ | ● | ○ | ○ | ● | |
| ARP spoofing | N/A | | ○ | | ○ | | ○ | ○ | |
| RKE cloning | N/A | | ○ | ○ | ○ | | | | |
| Malware | IDS, SNMP MIBs [55] | | ● | ● | ● | ● | ● | ○ | |
| | Trusted platform module [113] | | ● | ○ | ● | ● | ● | ○ | |

○ Targeted asset ● Protected asset

by Baker and Martinovic [77] for the use of a temporary Elliptic-Curve keypair until the session's ID key is produced and known to both parties, is a counter-MitM technique. By implementing the Elliptic-Curve keypair both the EV and the EV driver, as well as their credentials, are protected when accessing the CS by a MitM attack. These pre-charging negotiations can be protected by three-factor authentication protocol [98]. However, this protocol is mainly designed for the EVSE-CS internal communication.

OCPP security relies on lower-level protocols, such as on TLS, which in turn is not MitM-proof. Especially in the OCPP context, TLS is not sufficient, because [114]:

*(a)* it does not provide long-term authenticity or non-repudiation;
*(b)* it does not provide secure certificates validation;
*(c)* it introduces overhead; and
*(d)* it allows proxied data to be transmitted in clear text.

To prevent MitM attacks, a Back Propagation Neural Network (BPNN) scheme to be hosted by the CSMS was proposed in [99]. The BPNN can detect an MitM attack by analysing the charging/discharging requests and it can mitigate the attack with a delay- or a discard-request decision.

*2) **Packet replay attacks**:* One of the threats against the OCPP protocol is data disclosure. In data disclosure, an attacker can copy, read or replay private information about the EV and the EV driver. This information can be used by an adversary for financial benefits [36]. This information is usually obtained by eavesdropping or by a packet replay attack, where the attacker intercepts the communication between EV and EVSE through delays. This results to packet replays, which compromise the freshness of the OCPP messages. Replay attacks may occur when the exchanged messages are not encrypted and/or not authenticated.

To guard against these types of attacks, the Message Authentication Code (MAC) can be utilized [71] to secure the Controller Area Network (CAN) traffic between ECUs. However, MAC often does not fit into standard CAN data fields, which allow messages up to 8 bytes. Moreover, CAN messages are broadcast to all nodes without discretion, which means that they are vulnerable to eavesdropping or masquerading attacks. An intrusion detection system (IDS) [104],[105] proposed to train and recognize abnormal behavior can provide an alternative or a supplement to MAC.

This kind of attack can be mitigated with a variety of techniques. A three-factor authentication protocol can be used for the pre-charging negotiations between the EVSE and the CS [98]. An anomaly detection framework can be used with either machine learning techniques or a backpropagation neural network for detecting malicious OCPP traffic [109]. Specifically, in case there is the same OCPP version implementation on both the CS and the CSMS, the similarity between two consecutive requests and responses can be evaluated to show whether a pair of request/response is valid. So, the faulted CS can be identified and excluded from the charging process.

This type of cyber attack can also be mitigated with a TLS supported authentication of the OCPP messages,

which protects the communication between EV and EVSE. If TLS is implemented, the encryption of the messages with an asymmetric key will prevent session eavesdropping and hijacking [39]. However, TLS introduces unwanted overhead and additional costs, especially for the communication over cellular networks. So, a blockchain-based framework for EV reservation with a smart contract would be useful for the verification of the EV and CS through smart contracts. Thus, valid information and messages will not be replayed or eavesdropped [81]. In [106], such a blockchain-enabled EV charging system is proposed, with a focus on reducing the load on the grid, increasing user satisfaction and ensuring security. In these systems, EV and CS could authenticate themselves with freshly generated random numbers. However, such implementations lack the flexibility to adapt to the EV driver's changing preferences and to the dynamically changing grid conditions. Thus, for a protected PEV network, AI and Blockchain should complement each other. A distance bounding algorithm is proposed in [107] for the packet replay attacks on the EV-EVSE communication, which exploits the temporal delays created by the attack to the communication flows.

A packet replay attack could also be mitigated with the use of key-based authentication protocols in the pre-charging negotiations between the EVSE and the CS [98].

*3) **Denial-of-Service DoS/DDoS attacks**:* The main problem for the smart grid (SG) infrastructure is the development of optimal network design problems to address the existing gap between the security/resilience requirements and the provisioning of cost-effective SG communications. The security requirements are formulated in terms of monitoring the communications by the SG IDS, while the resilience requirements address the ability of the infrastructure to function in the presence of disturbances, e.g., failure or cyber attacks. The use of IDS engines and the optimal number of aggregators in the communication network among the smart grid components with a a one-layer aggregation-based machine-to-machine architecture can be studied based on three metrics: the energy consumption of each aggregator, the relay and aggregators cost, and the delay [115]. The use of TLS/SSL to secure communication with the combination of HTTPS and WebSocket Secure (WSS) is proposed in [109]. However, the implementation of TLS increases the overhead and since most CSs use the cellular network infrastructure, this results in an additional cost. Even with the use of a secure connection, basic requirements, such as end-to-end security and non-repudiation, are not always guaranteed.

Serious threats in the smart grid are the Denial-of-service (DoS) and the Distributed Denial-of-service (DDoS) attacks that severely threaten the availability of the Advanced Metering Infrastructure (AMI) network resources communication. These attacks target the CSMS, the CS and their communication links which use the OCPP protocol. The multiple EVs participating in the charging process can be used by an attacker to initiate a DoS or a DDoS attack by flooding the network with fake/unnecessary charging requests to reserve charging time slots. This type of attack overloads the charging schedules of the CS, and prevents the CS from serving benign EVs [39]. Moreover, the same attack can affect the EVSE communication channels by inhibiting the charging or by disrupting the grid services and possibly resulting to an unstable grid. This kind of cyber attack is a special disruption-type attack that comprises the deleting or dropping of messages. Some subcategories of DoS attacks like the grayhole attacks (i.e. selectively forwarding packets to the next hop) or the black hole attacks (i.e. dropping all messages), may severely affect the functionality of the OCPP protocol [64].

In the literature, the EV charging protocol should include processes for a scenario in which the EV reserves a charging time slot at a CS by sharing its private information with a trusted central management (i.e. the CSMS and the CS). CSMS and CS both make the reservation decisions for the EV [110]. Moreover, trusted third parties/central aggregators are assumed to handle the EV's private information. Thus, the security/privacy of an EV has been largely ignored. This private information can be either deleted or dropped by a malicious attacker who conducts a DoS attack. An important research challenge is how an individual EV could make an efficient CS selection in a decentralized manner, without leaking its private information neither to a CS nor to other network entities.

In order to address the above mentioned privacy and security problems, some works propose the use of AI in order to detect both random and faulted traffic using a single neural network [109]. An authentication protocol, which consists of a three-factor protocol for the pre-charging negotiations between the EVSE and the CS has been proposed [98]. Other works suggest the use of the blockchain-based distributed ledger technology (DLT) [111],[112]. More specifically, a DoS attack can be mitigated using a distributed blockchain-based efficient CS-selection protocol to ensure the security and the privacy of the EVs, the availability of the reserved time slots at the CS, and a high service quality. This blockchain-based EV charging solution enables the EVs to communicate with the CS without sharing their private information by either using a smart contract design or commitment schemes with binding and hiding properties in order for the EV to negotiate with the CS the best tariff option [81],[85]. The partially decentralized Consortium BC [108] can also be used. In this solution, only parts of the authorized nodes participate in the consensus, thus improving its efficiency and the transaction throughout. Consortium BC can also tackle the impersonation attacks, to which the OCPP protocol is also vulnerable, where an attacker pretends to be a CS or the CSMS.

*4) **ARP spoofing**:* In the Address Resolution Protocol (ARP) spoofing attack, the attacker manipulates the ARP messages to relate his MAC address to a legitimate IP address and receive data destined to the spoofed IP address. In the case of ARP spoofing, the PEV network information regarding the CS (i.e. location, availiability, status, charging profiles, schedules) and the EV (i.e. location, ID or other identification credentials) may be exposed [60]. This traffic integrity attack is usually followed by a MitM attack, where the MitM node is the one committing the ARP spoofing [71].

Although ARP spoofing is noted as a cyber attack against the PEV network elements, no OCPP-specific or PEV-specific countermeasures have been studied so far. ARP spoofing in PEV networks is currently addressed using conventional countermeasures of identifying an attack, relying on VPNs and static ARP, and on techniques such as the use of IDS, and packet filtering.

*5) **Remote Keyless Entry (RKE) cloning**:* The RKE systems support the keyless access to the EV, both to the driver and, unfortunately, to a potential attacker. An RKE-supported EV is accessed with a button click, which also activates a rolling code signal. However, the rolling code schemes can be eavesdropped, and, thus, the EV is vulnerable to RKE cloning [71]. An attack against the RKE systems is mainly targeting the EV's sensors. The more autonomous an EV is, the more dangerous an RKE cloning can become proven [95]. The OCPP operations are consequently affected by RKE cloning when a compromised EV enters the PEV network.

*6) **Malware attacks**:* An EV charging system relies on the safe and uninterruptible communication between the EV, the EVSE, the CS, and the CSMS [3]. All these entities can be attacked by malware that may lead to energy theft, data leakage or DoS. The former three entities are even more likely to get a malware installed, because of their physical exposure and the ease of access from an attacker to them. A malware attack is more likely to be executed by someone who has physical access or a privileged account in the PEV network (i.e. inside attack) [60].

The two cyber attacks described in [116] are based on malware. In the attack for bill reduction, the attacker misleads the customers of the PEV network, showing them a raised electricity price for the time slot he wants to charge and discouraging them from charging at the same time. This lowering of demand has the consequence of lowering the actual electricity price and the attacker manages to be served at a lower cost. In the attack for forming a peak energy load, the attacker identifies a peak time slot and misleads the customers of the PEV network showing them a low electricity price for the peak time slot encouraging them to charge at that time. This results in an energy load increase during the peak time slot and may lead to power overload.

To locate a malware within a PEV network, IDS and SNMP MIBs (IEC 62351-7) can be used [55]. According to [55], if the malware uses the EVSE as an entry point, the PEV intra-network communications will most probably be interrupted, whereas the charging service could continue using local default EVSE functions. Gharaibeh et al. [113] propose the use of TPM, which is a cryptographic dedicated module operating as a co-processor within a system under protection. TPM checks the system in every boot and prevents the booting if any change is detected.

## *E. Interdependent cyber-physical attacks*

The last group of attacks, the cyber-physical attacks (Figure 7), and the related countermeasures that are proposed in the literature are described in this Section. This group contains the attacks that have a cyber-physical impact on the assets of an EV charging system or the attack type is cyber-physical.

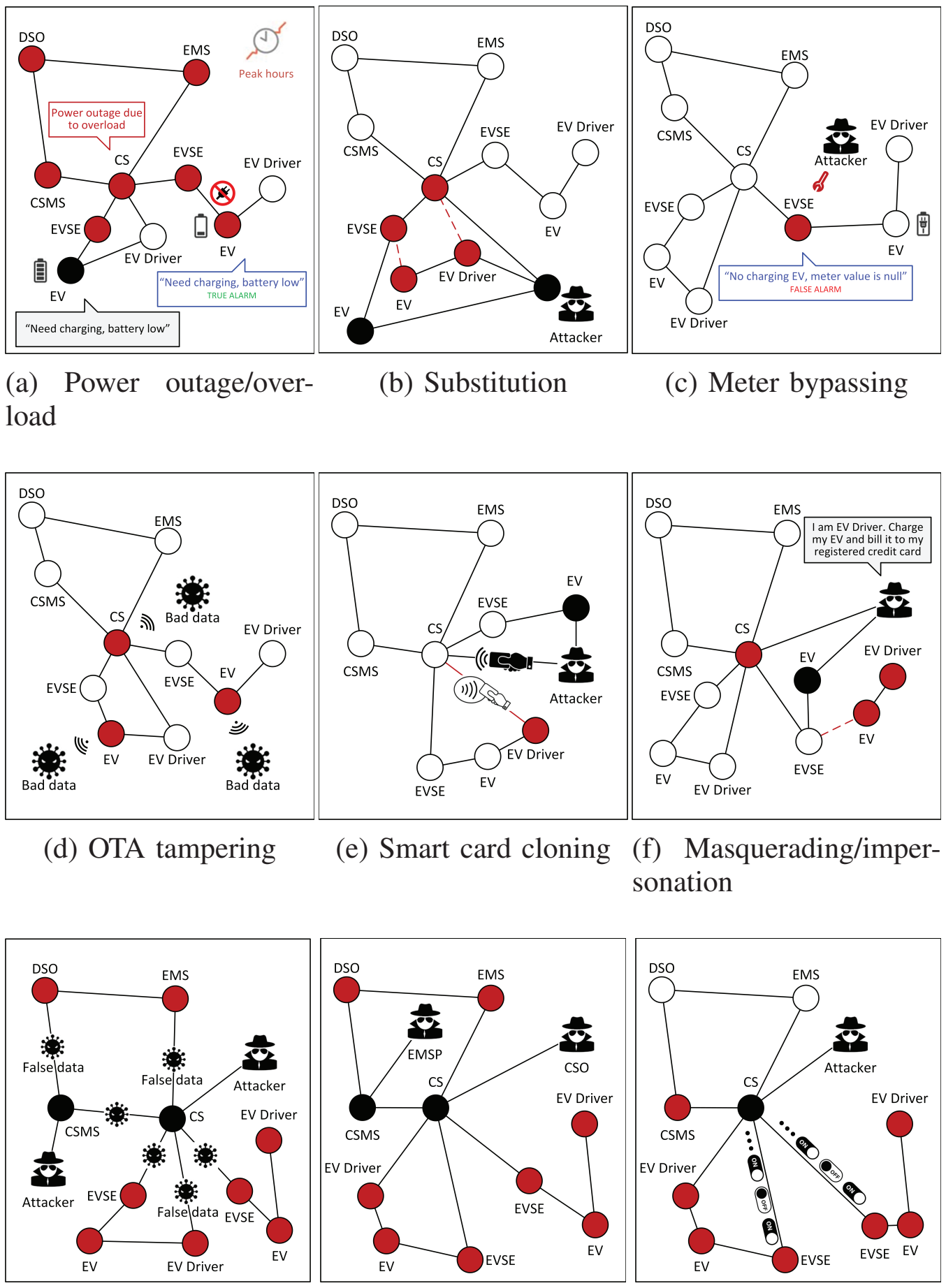

(a) Power outage/overload (b) Substitution (c) Meter bypassing

(d) OTA tampering (e) Smart card cloning (f) Masquerading/impersonation

(g) FDIA (h) Insider attack (i) Switching

Fig. 7. Cyber-physical attacks on PEV network

In Table VI, the related to every attack countermeasures are shown, followed by the assets of an EV charging system. The assets are either not marked if the attack does not affect them, marked as *Targeted asset* if the attack affects them and no countermeasure exists, or marked as *Protected asset* if the countermeasure is protecting the asset.

*1) **Power outage/overload**:* Both the power grid and the PEV network are co-independently vulnerable to cyber-physical attacks. The power outage is a cyber-physical attack that causes a charging service's unavailability and, hence, immobility issues. Power overloading is the other side of the same coin, a case where the attacker aims at increasing the energy load to disrupt the load balance on the local micro-grid, causing a blackout, and, thus, damaging the grid infrastructure [116]. Energy theft is a subcategory of these attacks; the attacker in this case has as a main target to pay less than the real value of the consumed energy [117]. Overloading is also considered to be a cyber-physical attack. The power utilities share publicly real-time information regarding the grid (e.g. outages logs or announcements, planned maintenance, upgrades and installations). This data weaponizes those planning to remotely deploy an attack against the grid [118].

*a) Protection of the micro-grid:* The increasing number of EVs and the forthcoming rise of energy needs call for enhanced grid infrastructures and flexible energy production, that combines the volatile and the renewable power resources. In this vein, Uhlig et al. [119] and Kubis et al. [120] propose an autonomous operation system, namely the InGO, for grid reinforcement. InGO consists of modules that divide the grid in low voltage and independently operating micro-grids.

In any case, power outages are a parameter that affects the charging system and can only be mitigated by a carefully designed power grid; it cannot be extinguished. The lack of any alternative solution, in terms of power for the CSs located in the outage reported area, would lead to the immobility of any EV and, hence, of the passengers in that area, as well. Amini, Mohammadi and Kar [121] propose a distributed decision-making algorithm that enables alternative energy resources to support the CSs without power supply and preserve the service's availability.

The impact on the power grid coming from the increased micro-grid load due to EV charging, especially during peak hours, was studied by the participants to the My Electric Avenue Project [122]. The outcome was that the load can be significantly controlled if every charging system implements a short intermission of less than 15 minutes during every charging lifecycle. This intermission, also known as supplemental reserve capacity, would lead to a lower SOC in some of the consuming EVs, which can be acceptable thanks to the constantly increasing battery autonomy, nowadays.

The protection of the OCPP-based charging site from the results of power outages was studied in [41]. The report presents the outcome of the aggregation of an EV fleet and of scattered CSs deployed at the Los Angeles Air Force base. The report proposes a hardware load separation (HLS) process for the EVSE and the local grid, and the supply of the EVSE via alternative power sources, such as via a storage battery system or via a diesel generator. The HLS would keep the charging service uninterrupted during the outage and the power grid relieved from additional start-up load on the power restoration phase.

In the same project, the PEV network includes an EMS, the Distributed Energy Resources Customer Adoption Model Optimizer (DER-CAM). DER-CAM combines data from both the fleet and a forecast system to manage the charging schedule for the site and to ensure the EV power autonomy in case of an outage, by enforcing a steadily high SOC for every EV. A similar EMS, the Independent System Operator (ISO), is proposed in [60] to manage the power load at peak hours, the charging cost and the schedule, and to support the service with backup power in the case of outages. The Outage and Restoration Management (ORM) is another example of an EMS designed to operate in the Field Area Network (FAN) of the smart grid's distribution tier, as described in the report of the programme ebalance-plus [56]. In the EU research programme named *SmartNet* [58] for the smart interaction schemes between the Transmission System Operator (TSO) and the DSO, and the EU research programme named *PlanGridEV* [123] for the grid planning and the operational principles of EV mass roll-out, an Outage Management System (OMS) is included in the architecture to not only identify and resolve outages, but also to keep logs of the grid's stability. Korba et al. [116] introduce a two level anomaly detection framework based on regression decision trees to leverage the regularity and the predictability of the energy consumption, to build reference consumption patterns for residential micro-grids, and, ultimately, to predict power overloading attacks. This solution affects the residential PEV networks.

*b) Protection of the power grid:* In an EV charging system, the EV connection to the grid should be verified before any charging process is initiated. During the charging process, any change on the connection or the existence of earth continuity resistance should pause or terminate the charging process, so that both the EV's battery and electric circuits, and the grid's power flow are not affected. In [124], the use of a physical control pilot conductor which controls the EV's charging load is proposed. The conductor and the control of the power flow with the use of a pilot signal is described in the ISO/IEC 15118 standard [31].

Some cyber-physical attacks exploit OCPP to affect the network infrastructure and, ultimately, the power grid. FDIA is an example of such an attack. The FDIA attacker uses the sensors operating within the PEV network and biases the values of the charging state of an EV or of a CS, without being easily detected. In [96], a method based on time interval observers to detect FDIA is proposed. In this case, the control center can identify the FDIA affected sensors by evaluating the time intervals in the sensor's data transmissions.

The Wireless Power Transfer (WPT) technology is a modern method for the EVs charging service provision within a PEV network. However, WPT suffers from vulnerabilities that could lead to a sudden and concurrent load or to a sharp load change caused by stationary energy (e.g. from the EV's battery). These vulnerabilities could affect the power grid's stability. The encryption method provided by the OCPP, when protecting the communication between the CS and the CSMS of the PEV network, is considered as a proper practice against most WPT vulnerabilities [125].

Some of the IDSs proposed in the literature deal with energy theft and grid overloading as well. Although treated similarly, the two attacks differ in several points such as in the attacker's operating mode, the detection delay, and the impact on the AMI. For example, a grid overloading cyber attack causes an immediate damage, therefore the detection delay in this case is more critical compared to the energy theft attacks. So, an energy theft detection system (ETDS) that can effectively detect energy theft attacks against AMI and can differentiate them from other cyber attacks is urgently needed. An ETDS would address two key issues. The first issue is the detection of malicious samples. The second issue is the sampling of the smart meters to achieve better load management and faster demand response. The higher the sampling rate, the higher the risk of revealing customers private information.

Jokar, Arianpoo and Leung [117] propose a consumption pattern-based energy theft detector (CPBETD) that employs novel techniques to overcome the problems associated with existing classification-based ETDSs. CPBETD utilizes

TABLE VI
CYBER-PHYSICAL ATTACKS ON OCPP-BASED EV CHARGING SYSTEM

| **Attacks** | **Countermeasures** | **Assets** | | | | | | | |
|---|---|---|---|---|---|---|---|---|---|
| | | **Driver** | **EV** | **EVSE** | **CS** | **EMS** | **CSMS** | **Data** | **Grid** |
| Power outage/ overload | Independent system operator [60] | | ● | ● | ● | ● | ● | ○ | ● |
| | Anomaly detection w/ regression decision trees [116] | | ● | ● | ● | ● | ● | ○ | ○ |
| | Consumption pattern-based energy theft detector [117] | | ○ | ○ | ● | ● | ○ | ○ | ● |
| | Autonomous grid operation system w/ firewall [119],[120] | | ○ | ○ | ○ | ○ | ○ | ○ | ● |
| | Charging distributed decision-making algorithm [121] | | ● | ● | ● | ● | ● | ○ | ○ |
| | Supplemental reserve capacity & limited SOC [122] | | ○ | ○ | ○ | ○ | ○ | ○ | ● |
| | Distributed energy resources model optimizer [41] | | ● | ● | ● | ● | ● | ○ | ● |
| | Outage and restoration management [56] | | ● | ● | ● | ● | ● | ○ | ● |
| | Outage management system [58],[123] | | ● | ● | ● | ● | ● | ○ | ● |
| | Physical control pilot conductor [124] | | ● | ● | ○ | ○ | ○ | ○ | ● |
| | Interval observers [96] | | ● | ● | ● | ○ | ● | ● | ● |
| | OCPP-encrypted CS-CSMS communication [125] | | ● | ● | ● | ○ | ● | ● | ● |
| | SOC-aware software-defined controller [126] | | ● | ● | ● | ○ | ● | ● | ● |
| | Smart meters [46] | | ● | ● | ● | ● | ● | ○ | ● |
| Substitution | Two-factor authentication [75] | ○ | ● | | | | | | |
| | Three-factor authentication protocol [98] | ● | ○ | | ● | | | ● | |
| | Multimodal multi-pass authentication [101] | ● | ● | | | | | | |
| Meter bypassing | Autonomous grid operation system w/ firewall [119],[120] | ○ | ○ | ○ | ● | | ● | ● | ● |
| | BC-based encryption, signatures [127] | ○ | ● | ● | ● | | ● | ● | ● |
| | EV localisation w/ road surface detectors [128] | ○ | ● | ● | ● | | ● | ● | ● |
| | Smart metering and peak shaving feature [129] | ○ | ● | ● | ● | | ○ | ○ | ● |
| | Role-based access control policies [130] | ○ | ● | ● | ● | | ○ | ● | ● |
| OTA updates tampering | AES 256 encryption [36] | | ○ | | ○ | | | ● | |
| | OCPP PnC mechanism [37] | | ● | | ○ | | | ● | |
| | Masked authenticated messaging module [131] | | ○ | | ○ | | | ● | |
| Smart card cloning | OCPP PnC mechanism [1] | ○ | ● | ● | ● | | | ● | |
| | Contactless banking cards [25] | ● | ● | ● | ● | | | ● | |
| | Location, time span, consumed energy monitoring [73] | ● | ● | ○ | ● | | | ● | |
| | Multimodal multi-pass authentication w/ smart card [101] | ● | ● | ● | ● | | | ● | |
| Masquerading/ impersonation | Cipher/hash-based message authentication code [63],[71] | ○ | ● | | ○ | | | ○ | |
| | BC payment [81],[108] | ● | ● | | ● | | | ● | |
| | Three-factor authentication protocol [98] | ● | ○ | | ● | | | ● | |
| | Multimodal multi-pass authentication [101] | ● | ● | | ○ | | | ○ | |
| | BC-enabled security architecture [102],[103],[132] | ● | ● | | ● | | | ● | |
| | Certificate revocation mechanism [113] | ● | ● | | ○ | | | ○ | |
| | Filtering using polynomials [133] | ○ | ● | | ● | | | ○ | |
| | Elliptic curve cryptography [134] | ○ | ● | | ● | | | ○ | |
| | Mutual server/EV authentication process [135] | ○ | ● | | ○ | | | ● | |
| FDIA | Interval observers [96] | | | | ○ | | ● | ● | ● |
| | Estimate network topology [136] | | | | ○ | | ● | ● | ○ |
| | Sequential change detection [137] | | | | ○ | | ● | ● | ○ |
| Insider | Role-based access control [100] | ● | ● | ● | ● | | ● | ● | |
| Switching | Back propagation neural network scheme [99] | ○ | ● | ● | ● | | ● | | ● |

○ Targeted asset ● Protected asset

appropriate clustering techniques and generates a synthetic attack dataset. This results to a robust against contamination attacks system, which achieves a high detection rate and a low False Positive Rate (FPR).

However, it should be noted that the EVs, depending on the grid operators (i.e. the DSOs), may be used as backup power generators in case of outage, because the EV batteries have the spinning reserve capacity. The batteries also support black start services (i.e. the process of power restoration relying only on distributed energy and not on external sources) and the EVs may be used as a key element during the restoration phase after an outage [138],[139]. These features, if evaluated by a cyber-physical attack, may disturb the power grid with *(a)* a sudden surge in demand (i.e. many EVs appear to request charging at the same time), *(b)* a sudden surge in supply (i.e. many EVs discharge power from their batteries to the grid simultaneously), or *(c)* a switching attack (i.e. multiple charging and discharging activities in a short time period to destabilize the grid) [3].

In this vein, Li et al. [126] propose the implementation of a software-defined controller, the SD-V2G, which is constantly evaluating the EV's SOC-aware flow and facilitates the transition from state to state (i.e. the charging, suspending and discharging states) of the EV. This two-way flow of data and power (also referred to as Vehicle-to-Home (V2H), inspired by the residential CSs) is proposed to use smart meters, which detect outages on behalf of the micro-grid or of the power grid operators [46].

*2) **Substitution attacks**:* The identification and authentication over the power grid processes may be physically bypassed by an EV connected in the place of an already authorized EV. These attacks, called substitution attacks, are a cyber-physical variation of the MitM attacks. The EV's authentication process, especially when the EV is wirelessly connected to the EV charging system, may expose the EV's identification credentials. The scenarios of a successful substitution attack include the activation of the charging and billing processes for a falsely authenticated EV based on *(a)* the driver's credentials alone or, *(b)* the credentials of a different EV [75]. The substitution attacks are motivated by the stealing of valid EV credentials on behalf of an unidentified EV and may escalate to EV or EVSE tampering, expressed by the stealing of the ECU which hosts the EV credentials [31] or by the transferring of the charging cable from a legitimate EV to another EV.

In [75], a two-factor cyber-physical authentication process is proposed where the physical connectivity parameter is, also, evaluated alongside with the EV's digital ID. In the same scope, Vaidya and Mouftah [101] propose a Multimodal Multi-pass Authentication (MMA) mechanism which identifies and authenticates both the EV and the driver, while the respective credentials are obtained from multiple paths. MMA has been also tested using contract certificates in compliance to the ISO/IEC 15118 standard. Irshad et al. [98] propose a three-factor authentication protocol for the pre-charging negotiations between the EV and the CS. The EV is authenticated based on the EV driver's credentials, which include a bio-metric factor. The protocol is supplementary to the OCPP and is proven resistant to substitution, man-in-the-middle, impersonation, and denial-of-service attacks.

*3) **Meter bypassing**:* The meter bypassing cyber-physical attack takes place on the EVSE [60] and, as it results on altering the billing process, it also affects the EV driver. Although meter bypassing needs physical access and manipulation of the EVSE hardware, it can be cyber-physically conducted by exploiting the smart metering features and applications. The meter bypassing may affect *(a)* the EV driver in terms of service and billing, *(b)* the EV, which may not be charged, and *(c)* the power grid, that may be overloaded or under-priced.

OCPP supports smart metering for the communication between the CPO and the power grid operators (e.g. the DSO) to protect the metering and billing processes. The security of the OCPP-based communication between power suppliers and the back-end billing system is a concern of the Penta Security, the South Korean company that developed the AutoCrypt V2G [127]. AutoCrypt V2G is a blockchain-based suite which provides data encryption and digital signatures for the authentication and for the authorization of the entities participating in a charging system. AutoCrypt V2G also supports the PnC feature of OCPP.

The smart metering for the publicly accessed charging systems is usually based on the localisation of an EV for billing. For that reason, a common practice is the availability of one CS per parking bay. The integrity of the billing process relies on two assumptions: *(a)* that every CS-supported bay will be occupied only by EVs, and *(b)* that every EV will connect to the corresponding CS and not to a nearby sited CS. Both conditions, however, are not ensured in the case when the driver moves the EV charging cable from one CS to another. Nürnberg and Iwan [128] propose the use of magnetic, inductive, or electromagnetic detectors placed under the road surface that provide real-time localisation data for all EVs in the parking area.

Meter bypassing is a security issue not only for the EV charging system and the provided service, but, also, for the power grid and the provider's back-end systems. The need for power is massively increasing as every fossil-fueled vehicle is replaced by an EV. Smart metering enables options, such as peak shaving [140]. Peak-shaving allows the DSO or any operator from the power grid's control side to reduce energy consumption in peak hours for the protection of the grid. This is proposed in the case of the Motown [129], an open-source platform proposed by E-laad, Alliander Mobility Services (AMS or Allego), iHomer, and NewMotion.

The constantly increasing number of EVs and the forthcoming rise of energy needs call for robust grid infrastructures and flexible energy production that combine the renewable power resources. In this vein, Uhlig et al. [119] and Kubis et al. [120] propose an autonomous operation system, namely the InGO, for grid reinforcement. InGO consists of modules that divide the grid in low voltage micro-grids which are independently managed. In the case of an OCPP-based EV charging system, InGO intervenes between the CS and the CSMS communication and protects the data exchanged with the InGO firewall, resulting in the safe processing of the smart meter data.

The research group of the *NOBEL GRID* EU programme proposes to harden the smart meters with the use of a smart meter extension module (SMX) [130]. SMX applies role-based access control (RBAC) policies to restrict unauthorized access to the smart meters.

*4) **Over-The-Air (OTA) updates tampering**:* The EV and the CS contain a hardware complex, where each device has a different firmware. Considering the firmware constant changes, the cost of recalls and the time cost of calling an EV in for a firmware update, the update Over-The-Air (OTA) approach is an appealing alternative. OTA is facilitated by the almost uninterruptible network connection of modern EVs. Nevertheless, OTA updates are both a security risk and a security countermeasure. The OTA tampering of a firmware or of a software can expose vital EV operations to remote attacks (i.e. control override attacks or Remote Control Execution (RCE) attacks [106],[141], and, at the same time, an EV with an outdated firmware or software may be vulnerable to the latest cyber attacks [71]. The same problems are faced by the CS.

Whereas OTA firmware updating and patching is a convenient way to bypass recalls and out-of-service periods for an EV, firmware is left exposed to tampering while being transmitted to the EV. Additionally, the OTA option is not always supported by the EV, especially by older models. Buschlinger, Springer and Zhdanova [37] propose the updates to be done via the PnC connection of the EV to the CS. The transmission of the firmware will be done over the charging cable and the firmware will be tamper-proof. According to [37], the mean time of a charging process, i.e. 30 minutes, is enough for a simultaneous update, offered as a Value-Added Service (VAS) in the charging site. Such services are already under study and development [142].

An anti-tampering solution for the protection of the firmware or the software transmitted as an OTA update is the use of the Masked Authenticated Messaging (MAM) module [131]. MAM allows the seeding of data via channels. The owner of the seeded data and of the channel may change the access rights to public, private or restricted and, thus, it will be feasible to control who will be able to access the seeded data and with what rights.

A list of actions is proposed to the EV manufacturers to minimize the risk of remote access, which is a direct result of OTA updates tampering [36]. The list includes the following countermeasures:

*(a)* authorization and authentication process executed prior to the update process;
*(b)* encryption of the update's files, the EV data, and the inter-ECU messages; and
*(c)* restore/revert option in case of OTA update failure or tampering.

The encryption should be done using the Advanced Encryption Standard (AES) 256 algorithm or other acceptable standardized alternatives.

*5) **Smart card cloning**:* The EV driver is usually identified within the PEV network by the Unique Identifier (UID) of an Radio Frequency Identification (RFID) smart card [73]. The smart cards are also used as micro-processors and storage devices to secure the driver's data from tampering [72],[74]. The cards are issued by the managing authority of the PEV network (nowadays the same authority has the role of the CPO based on the contract with the EV driver). However, the smart cards may be cloned, which is an OCPP impersonation attack variation. The protocol may be severely affected by such an attack, if the attacker accesses the OCPP parent idTag, under which a group of tokens can be manipulated [64].

The smart card cloning attacks are already popular despite the current low prices of charging because they are easy to deploy and also they have an immediate economic benefit when reselling the cloned cards or using the cloned cards for a free-of-charge service. However, there are techniques such as the monitoring of the location, the time span, and the consumed quantity of energy that can help identify a cloned card in a PEV network [73]. For this reason, they are characterized as attacks of medium-high impact, high feasibility, and low detectability.

The tag ID of an RFID smart card, which is read at a CS, combined with the EV driver's login credentials to the charging scheduling/billing application are used for the authorization of a charging service in the Multimodal and Multi-pass Authentication Scheme using smart cards (MMA-SC) [101]. The service is activated only for the CS where the card was read.

Aubel and Poll [25] propose the replacement of the RFID smart cards with the use of contactless banking cards, supporting the EMV standard (EMV was originally created and named after the three companies Europay, Mastercard, and Visa). According to EMV, it is possible to authenticate the card with the stronger security-wise asymmetric cryptography. Alignment with the EMV means that the card readers embedded on the CS should also support the EMV authentication method.

The latest OCPP 2.0, supports the PnC mechanism for the identification and authentication of the EV [1] upon connecting to the CS, to tackle the EV driver's exposure via the use of a smart card.

*6) **Masquerading and impersonation attacks**:* A PEV network may suffer by masquerading attacks (e.g. illegitimate EV charging with another EV's credentials) or by the slightly different impersonation attacks (e.g. an EV driver charging his EV and billing another EV driver) [100]. The main entry points for both these attacks are the EV, the EV driver (e.g. the credentials of his credit card, the access account for the reservation/billing application), and the CS [24],[39],[62]. The masquerading attacks are mostly used to disrupt and to affect the integrity of the charging service, while the impersonation attacks are used by the malicious party so that the service's control will be taken over and the service's availability will be disrupted [143].

A generic but valid countermeasure to prevent masquerading attacks is to use an identification method based on certificates issued by trusted authorities. In [113], the certificate revocation mechanism is proposed to ensure the validity of the certificates against post-issuance masquerading. The Multimodal Multi-pass Authentication (MMA) mechanism [101] is another case where the protection of the EV credentials is achieved. The aforementioned three-factor authentication protocol protects the OCPP against EV driver's impersonation attacks [98].

Masquerading attacks may occur when messages exchanged between ECUs are not encrypted and authenticated. Since the Controller Area Network (CAN) messages are broadcasted to all nodes without discretion, they are vulnerable to masquerading attacks. Since CAN frames are not encrypted, the carried data can be accessed by the attacker who will be able to locate system entry points. To guard against this type of attacks, the Hash-based Message Authentication Code (HMAC) can be utilized [71] for securing the CAN traffic between ECUs. The cipher-based variation of MAC (CMAC) [63] secures the intra-ECU communication utilizing the AES cipher and shared keys between communicating ECUs. In [133], an en-route filtering scheme using polynomials instead of MAC is used to protect the nodes. The encryption of the V2G communication this time is treated by a proposed ECC-based protocol (ECC stands for Elliptic Curve Cryptography) [134].

As a mitigation approach, Danish et al. [81] implement the payment functionality in blockchain to make sure no false information can be injected in the system and that the EV will securely pay only for the provided service. As the EV and the CS are making reservations and charging payments on a blockchain network, no malicious entity can spoof their identity during the charging process. Evaluations show that the proposed BlockEV is scalable with significantly low blockchain transaction and storage overheads.

However, the deployment of smart contracts in blockchain lacks the flexibility to adapt to the dynamic charging behavior of the EV drivers and grid conditions. In the consortium blockchain solution [108], only parts of the authorized nodes participate in the consensus and, therefore, impersonation attacks against the OCPP protocol are contained. The consortium blockchain solution is only registering nodes verified by multiple key pairs. Moreover, the keys are different for every transaction of the node. A similar approach is presented with the EV power trading decentralized architecture based on the consortium blockchain [102],[103], where an EV impersonation attack is prevented using a private key (PK) and a secret key (SK) for every legitimate digital signature and an improved Krill Herd (KH) algorithm. Similarly, the blockchain-enabled security architecture for EVs based on cloud and edge computing [132] uses blockchain-based data and energy coins, while data frequency and energy contribution amount are evaluated for the security level determination.

As discussed earlier, the OCPP PnC feature is an anti-tampering tool for the EV and its hardware. However, if an attacker manages to get possession of the PnC credentials, he may deploy an EV impersonation attack and *(a)* commit fraud by billing another legitimate costumer, *(b)* get free access to value-added services, or *(c)* request renewed PnC credentials and get falsely validated. The generation and storage of the credentials within the EV can protect the PnC credentials [89].

A more generic but effective solution, the embedding of PUF on the exposed EV and CS [85],[86],[87], will allow the identification of an impersonator or a masqueraded device in the PEV. It is notable that ISO 15118-2 considers the server-side authentication as mandatory, whereas it acknowledges that a mutual server- and EV-side authentication process would prevent impersonation attacks [135].

*7)* ***False Data Injection Attacks (FDIA):*** FDIAs can be performed by compromising the communication channels. Typically, an FDIA attack is conducted by hacking the components of the overall charging architecture using the OCPP protocol, or by accessing and manipulating the database of a control centre, by sitting at the CSMS component.

FDIA in the setting discussed in this survey aims to compromise the meter measurements. Remember that the system, including the communications over the V2G network, is largely regulated based on the EV state's estimates. The attacks target precisely these estimates, and if effective they can lead to wrong decisions, e.g., about the charge level, the timings, or even the market prices. There are effective methods for detecting bad data in networked systems, such as the least squares (LS) estimator-based detectors, where binary decision diagrams (BDDs) are used to detect bad data due to random noise and faults, but not for detecting FDIA. Bobba et al. [144] aim to constrain attackers to only a limited number of meters/nodes in order to protect the entire network. Sou, Sandberg and Johansson [136] aim to tackle attacks which commence after the perpetrator has collected offline or electricity consumption data and use that to visualise the topology of the network from such incomplete knowledge.

Kurt, Yilmaz, and Wang [137] used a sequential change detection approach as the quickest way to detect an FDIA. The idea builds on the fact that LS estimator-based methods depend only on present measurements, while their approach is to adopt a state-space model that enables the use of both past and present values, so that accuracy is improved. In [96], the design of interval observers series considering the bounds of internal states is proposed. This involves the use of modelling errors and disturbances to estimate the interval states of the grid physical system on which the OCPP protocol operates. FDIA detection can be achieved by using the interval residuals of the observers. The measurement data of the corresponding sensor (e.g., meter) is used as an input to the interval observer. A logic localisation judgement matrix is constructed to localise the sensor in which an FDIA has taken place. Therefore, interval residues observers and a logic-based localisation matrix for FDIA detection are proposed [96]. This approach is demonstrated on the IEEE 36-bus grid and can be adapted in a straightforward manner to smart grids/networks running the OCPP protocol.

*8)* ***Insider attacks:*** The CS/EVSE tampering attacks are a usual expression of insider attacks; an insider is a person considered to be trustworthy to the EV charging operators and is therefore given access privileges. An insider can be the actor in an attack for energy theft [64]. There are also some 'hybrid' roles, such as the EMSP and the CSO, which are not always fulfilled by a physical entity, but, sometimes, they are fulfilled by a person or a device that has another role already. The existence of these roles is a security gap that can be exploited by an insider. In a study regarding the end-to-end security in smart EV charging, the communication between the CSO, the LC and the CS is mostly targeted by an insider attacker, because of its simple execution [73].

IEC 62351-8 Role-Based Access Control (RBAC) [100] introduces roles and inherited permissions and restricts access to attackers, even if they are insiders. RBAC can protect the aforementioned devices, entities, and data and, consequently, OCPP communication.

*9)* ***Switching attacks:*** The switching attacks are attacks with a cyber-physical impact. They use the OCPP *setChargingProfile* or the *startTransactions* requests on the CS. When the CS is compromised, these OCPP messages allow the attacker to install malicious charging profiles. With these profiles in action and the control of the CS, the attacker can activate multiple charging/discharging activities in short intervals and, ultimately, destabilize the service and the grid [3].

The switching attacks can be mitigated using the Back

Propagation Neural Network (BPNN) scheme [99]. The BPNN algorithm assists in the suspicious requests detection by analysing the charging/discharging requests. If such requests are detected, the algorithm discards the requests or inserts delays in their execution time to limit the impact of the attack.

Other countermeasures to prevent switching attacks from succeeding, are the following [3]:

*(a)* Strict CS access control policy for the allowed data traffic, restricted access to physical or network ports, strong credentials and authentication methods.
*(b)* Anomaly detection engine to monitor the EV schedules.
*(c)* EV driver's approval/disapproval for any charging schedule update.
*(d)* Contingency plan for the power grid.
*(e)* Anomaly detection engine to monitor the smart meters' data streams.
*(f)* Standardization within the current charging infrastructure.

## V. Privacy Issues & Detection/Deflection Mechanisms

Considering the architecture and the underlying infrastructure for charging EVs, privacy data can be leaked in various places. Private data includes personal information of EV owners/customers, including patterns of energy usage, types of EVs, movement pattern of EVs and owner(s). Private data of the EV itself includes the timings of when an EV needs to be charged, at what location(s), owner/user data, payment details, energy usage, and frequency of charging. When the privacy of such data is compromised an adversary can know when the owner is travelling, and maybe even in which direction. If the EV ID is stolen the adversary can charge other EVs in its place and monetarily affect the user(s) of the compromised EV. On these grounds, detection and deflection mechanisms are of great importance for minimizing the impact of such illegitimate actions.

TABLE VII
Privacy Issues and Detection/Deflection Mechanisms

| | Reference | Method |
|---|---|---|
| **Privacy preservation/ Authentication/ Authorization** | [145],[146],[147] | Fully/partially/restrictive blind signature |
| | [148],[149] | Master-/Manager-based group signature |
| | [150] | Ring signature |
| | [151] | Link Aggregation Groups (LAGs) secret sharing |
| | [152],[153],[154] | Homomorphic/symmetric/asymmetric encryption |
| | [74] | Third-party anonymity |
| | [155] | Anonymity networks |
| | [156] | EV mobility in destributed/centralized V2G |
| | [50],[157],[158] | Plug-and-Charge (PnC) EV mode |
| | [101] | Contract certificate (CCert) and V2G PKI |
| **Detection/ Deflection** | [159],[160] | Distributed IDS in AMI network |
| | [117] | Energy Theft Detection Systems (ETDS) |
| | [161] | Honeypots in AMI network |
| | [115],[159],[160],[161] | Low energy consumption |
| | [162],[163] | Anti-jamming/spoofing |
| | [104],[105] | Machine learning abnormal behavior detection |

Regarding privacy issues, EVs have to provide their IDs, plug into charging spots, connect to LAGs, and communicate with service providers. This leads to potentially exposed users private data, such as the location of charging or discharging [164]. The location privacy of the Judging Authority (JA) for payment and billing is also discussed in [165].

Battery management also poses a potential threat to privacy data as battery information can provide additional information for the adversary which can be used to build the EV owner's mobility profile. Furthermore, it is much easier to attack V2G networks rather than the traditional grid due to the two-way communication taking place between EV and the grid [166].

Data management in V2G networks includes data collection, aggregation, storage, and publication. Attackers may show a great interest on hacking the database or the storage systems. Furthermore, various malicious software and attacks are aiming at database or storage systems.

Billing is also done differently in V2G networks than in the traditional power grid. Although a credit card can satisfy most of the needs of V2G networks, it is not a good way to protect privacy since in many financial applications, even though credit numbers are encrypted, some information related to credit cards is not encrypted [167].

V2G networks may employ OCPP everywhere or they may use ISO/IEC 15118 as the charging protocol between the EVs and the charging spots, adopt IEC 61850 for the communication between charging spots and energy providers, and use OCPP as the communication protocol between the charging spots and the mobility operators. All of these protocols have their own specific vulnerabilities. Potential attackers may make use of the loopholes in these protocols for malicious purposes.

Based on the method used, Table VII provides a classification of the literature findings regarding the privacy preservation techniques, the authentication/authorization mechanisms, and the detection and deflection mechanisms that apply to OCPP-based EV charging systems. These techniques, mechanisms, and tools are discussed in the following Sections.

### A. *Privacy preservation techniques*

Some privacy preservation techniques in V2G networks have been developed recently. A number of approaches adopt the notion of a *signature* which is key to authentication schemes; these are further discussed in Section V-B. There are schemes that use a fully blind signature [145], a partially blind signature [146], and a restrictive blind signature [147]. An overview of blind signatures in the context of V2G networks is given in [62]. The method of blind signing is suitable for privacy-related schemes such as a PEV network, where digital payments take place in public areas, since the data is disguised before signed and, therefore, can be publicly verified with limited exposure.

There are also schemes that propose the use of a *group signature* which allows one member of a group to anonymously sign a message on behalf of the group. These are classified into master-based [148] and manager-based group signature schemes [149]. In PEV networks, a group signature scheme would allow the CS to anonymously authenticate and dynamically manage the connected EVs, relieving the CSMS from the corresponding computational cost [168].

Additionally, schemes based on *ring signature* have also proposed. This method allows each member of a group to sign a message without revealing the member's identity [150]. In contrast to the group signature, the group in a ring signature is formed in an ad-hoc basis. In [150], a ring signature that aims to protect the privacy of an EV when it plays different roles in V2G networks is proposed; an EV can be a customer when it charges from the grid while it can be a generator when it discharges back to the grid or supplies electricity to other EVs, and a storage place when it does not need charging or discharging.

*Secret sharing* techniques have been proposed to address anonymous data aggregation. In [151], a secret is broken down to several parts held by different members. To restore the secret, all or some parts are required. The plug-in time, the current level of battery charge, and the amount of recharged electricity are three types of data that are typically held in the EV charging context by different LAGs. To compromise the EV data, all LAGs need to be compromised at once.

*Homomorphic encryption* is another approach to address data aggregation. It comes in different flavours, including fully or partially homomorphic encryption [152], and symmetric [153] or asymmetric encryption [154].

Other approaches include *third-party anonymity* [74], which employs a third-party secure and trusted device to hold sensitive information of other (non-secure) devices (e.g. the EVs in a PEV network), and *anonymity networks* [155], which are communication networks that are designed with a Certificate Centre and a Public Key Infrastructure to conceal network layer IDs in V2G networks.

The aforementioned techniques introduce an additional computational effort on the architectural elements of a PEV network, while they support the data privacy. Depending on the infrastructure and the resources available in each public-access PEV charging site, residential charging site and so forth, the most suitable scheme for privacy reservation is applied.

### B. Authentication & authorization mechanisms

Authentication and authorization are key issues for the privacy preservation of the EV charging process [39] and of the EV and the EV driver credentials. For instance, an authentication scheme considering the mobility of the EVs in distributed, as well as in centralized, V2G networks is proposed in [156].

Regarding the EVs authentication, the ISO/IEC 15118 standard [50] includes a provision for the secure EV connection to the charging facilities. This comes in two identification modes, namely the *PnC mode* and the *External Identification Mode*. However, it has been argued that the standard has some drawbacks [157],[158]. For example, the PnC mode can authenticate the legitimate EV only, while the External Identification Mode can authenticate the legitimate EV user only. This means that an unauthorized EV may be charged with a valid EV user's smart card. It also means that an EV with an authorized digital certificate may be allowed to charge even if the EV owner is not a valid user.

In short, the standard defines a certificate-based method for authentication/authorisation purposes, which is uni-modal and single path, and, therefore, provides low levels of security. Adversaries can exploit this to launch active attacks [157] such as MitM and substitution attacks [169].

Recent proposals to mitigate such attacks involve authentication mechanisms which deploy multiple modes of credentials obtained from multiple paths, before the CS allows the EV owner to initiate the charging process. A Multimodal and Multi-pass Authentication Scheme using Contract Certificate (MMA-CC) has been recently proposed in [101]. This scheme has two phases. First, a bootstrapping phase where a contract certificate (CCert) is issued to the EV to support secure and reliable updates from an Original Equipment Manufacturer (OEM) [37]. Then, and only after the CCert is obtained, the charging process continues to the MMA-CC operational phase, where the CCert is sent for validation. The charging can begin only after an authentication response on the legitimacy of the credentials is received.

A smart card based scheme, called Multimodal and Multi-pass Authentication Scheme using Smart Card (MMA-SC), has also been proposed by the same researchers. MMA-SC deploys RFID or Near-field communication (NFC) smart card technologies and targets EVs which do not have digital certificates neither support V2G PKI [101]. This opens up the scheme to classic smart card cloning attacks, but the multimodal aspect of the scheme ensures that these are not effective.

Moreover, a legitimate contract certificate and valid user credentials are required in the MMA-CC scheme, while a legitimate smart card tag ID and valid user credentials are also required in the MMA-SC for a successful EV charging process; therefore, MitM attacks can be mitigated. Regarding the substitution attacks in both schemes, the EV charging process does not proceed until multimodal and multi-pass authentication, i.e., CCert or the smart card and the user credentials are validated, thus a substitution attack is not possible.

### C. Detection and deflection mechanisms

The EV smart charging procedures inherit a multitude of security concerns and vulnerabilities from all the participating actors. These vulnerabilities are associated with the actors themselves, i.e. the EVs (e.g. impersonate attacks, physical damages), the exchanged messages between the CSs (e.g. privacy, tampering), and the communication medium used (e.g. DoS, DDoS, MitM, RF jamming, eavesdropping) [39]. The threats associated with the OCPP architecture and the wireless communication may additionally relate to [64]:

- *(a)* Disclosure, which corresponds to the illicit reading and/or copying of information.
- *(b)* Distortion, any (fake) data insertion, spoofing or modification action against data, processes or configurations.
- *(c)* Disruption, that comprises the deleting or dropping of messages, processes or actions.

To address these security concerns in an OCPP architecture, a distributed IDS can be deployed in different sensor locations

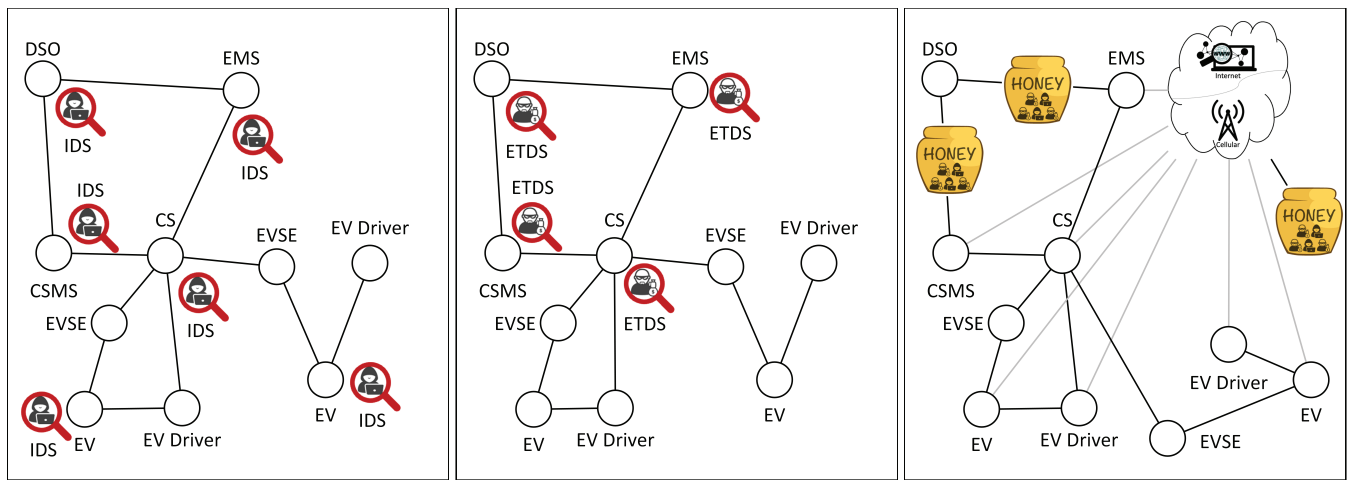

(a) Distributed IDS (b) ETDS (c) Honeypots

Fig. 8. Detection and deflection mechanisms on PEV networks

of the network [159] (Figure 8(a)). The distributed IDS approach complies with PEV networks requirements for [64]:

*(a)* lightweight detection mechanisms based on knowledge of the environment like, for example, restricted IDSs, and

*(b)* lightweight trust-based systems to make sure that the information received by a node is reliable and the network connectivity of the nodes is uninterruptible.

In [159], the information required for detecting attacks against distributed power metering infrastructures, such as AMI and PEV networks, is classified as:

*(a)* *System information*: health reports from meters, battery consumption, power overloading, energy theft, software integrity of AMI devices and clock synchronization.

*(b)* *Network information*: collision rate, packet loss, node response time, traffic rate, health and integrity of transmitting messages, associations between physical addresses, and node identity.

Attacks involving the *System information* category affect the AMI, which provides a two-way communication network between smart meters and utility systems; the attacker offers interactive services for managing the billing and the electricity consumption. However, the interconnection of the smart grid distributed elements also introduces new vectors for cyber attacks. Such cyber attacks concern either the power overloading or the energy theft attack and they can be mitigated with an energy theft detection system (ETDS) that can effectively detect energy theft attacks against AMI and differentiate them from other cyber attacks (Figure 8(b)). The ETDSs use either classification-based or cluster-based solutions for designing a pattern-based anomaly detection framework. This type of continuous multi-level monitoring of the energy consumption load allows for an efficient and early detection of such cyber attacks. An example of the use of an ETDS is the consumption pattern-based energy theft detector (CPBETD) [117], which is robust against contamination attacks and non malicious changes in consumption patterns, achieves a high detection rate and has a low False Positive Rate (FPR).

In the *Network information* related attacks, the main problem for the Smard Grid (SG) infrastructure is the network's design, which lacks the ability to address the existing gap between security/resilience requirements and the provisioning of cost-effective SG communications. Such a network needs to support the monitoring of all communications by an IDS, and to provide uninterrupted functionality even in the case of failure or under cyber attack. The placement of IDS engines and the optimal number of aggregators should be done with a consideration of the energy consumption of each aggregator, the relay and the aggregators cost, and the resulting delays [115].

A major issue is the location of the IDS in the network and the trade-off between cost and security. In [161], the use of honeypots in the AMI network as a decoy to detect DDoS/DoS and gather attack's information is proposed (Figure 8(c)). The design of a resilient SG communication infrastructure, where IDS engines are spread across to ensure the monitoring of flows at a minimum cost is shown in [160]. The work explores a column-generation model-based approach to address the detected issues in short computational time. Similarly, a cost model-based framework to aid utilities in the provisioning of IDS is proposed in [159]. The framework leverages the output of risk assessment methodologies, which represent the input to a decision-assistance model. The model can be used to analyze the trade-off between cost and benefits of installing IDSs in different locations. This work also integrates the communication network (connectivity matrix) into the decision process, and identifies the optimal positioning of IDS.

Lastly, there are defense mechanisms that can protect the charging of EVs and detect both jamming and spoofing attacks. A novel cross-layer IDS for detecting spoofing attacks against the wireless communication of connected EVs is proposed in [162]. Gai et al. [163] also introduce a new attack mechanism that uses both jamming and spoofing to intervene in the normal wireless communications of the smart grid adopting cognitive radio networking approaches.

All the aforementioned IDSs have the potential to detect attacks which originate from the AMI network and move towards MAC or PHY layer attacks, and end-to-end application layer attacks between P2P AMI nodes. The main metrics that are used for the detection are parameters, such as the Signal to Interference and Noise Ratio (SINR), the Packet Delivery Ratio (PDR), and the end-to-end delays.

Masquerading, eavesdropping, injection and replay attacks can occur when messages between the ECUs are not encrypted and authenticated. To guard against these types of attacks, MAC can be utilized [71] to secure the Controller Area Network (CAN) traffic between ECUs. However, MAC often does not fit into the standard CAN data fields. Moreover, CAN messages are broadcast to all nodes without discretion. In [104],[105], an IDS which incorporates machine learning in order to train and recognize abnormal behavior, is proposed as an alternative or a supplement to MAC.

Concluding, an open issue regarding the OCPP security is the integration of the network-related IDSs and the system architecture-related IDSs with the main objective of a cross-layer IDS design capable of detecting every kind of attack, whether it concerns the application layer or those concerning the lower layers. So, finding the best trade-off for a scalable and comprehensive cross-layer IDS is key to investing in the right technology and deploying sensors at optimal locations.

## VI. Open Issues and Future Directions

### A. Lessons learned digest

OCPP version 2.0 is the first protocol's version that includes implicit security blocks. Although security was also a challenge for the previous version developers (i.e. version 1.6), the issuing of server certificates using the TLS protocol is not optional in OCPP 2.0 as it was in the previous two versions. To implement a more secure protocol profile, the TLS is mandatory and supports the mutual client-server authentication [25]. However, till today, previous versions of the protocol are still used; for example, many implementations rely on the older versions, even on version 1.5. Therefore, the security related research on the OCPP is very important. Moreover, even for OCPP 2.0 where TLS is mandatory, the version of TLS used is also a factor for the protocol security level.

In the literature, several types of attacks are studied, and, in most cases, countermeasures are proposed. We provided an exhaustive analysis of the security related research studies, grouped in *(a)* physical attacks, *(b)* cyber attacks, and *(c)* cyber-physical attacks. The studies referred in this survey discuss the security issues across all versions of the OCPP.

It is interesting that most research outcomes on the OCPP security are related to the tampering, the Man-in-the-Middle attacks, the power outage/overload, and the masquerading/impersonation attacks. This can be explained by the following PEV network characteristics, respectively:

- Important devices/assets, e.g. the CS, are exposed in public access.
- Communication messages exchanged are, in many cases, not encrypted and carry important/sensitive information, e.g. the EV driver's credit card data.
- Service availability is important to the EVs, the consumers, and the power grid, in terms of profit.
- Information regarding the billing/payment process may be exposed to an attacker and directly converted to profit.

The existing studies have another interesting characteristic: the countermeasures proposed in the literature do not cover the protection of the affected assets in every attack. In the group of physical attacks, specifically regarding tampering that affects all assets except for the grid, fifteen (15) countermeasures are proposed, two (2) of which are protocol features, none considers the EMS protection and only two (2) of them consider the CSMS protection. In the same group of attacks, no countermeasure against the sensor attacks is proposed.

In the group of cyber attacks, the MitM is on the spotlight, being the most common and dangerous type of attack. Out of the nine (9) proposed countermeasures, while none is focusing on the EMS protection. Additionally, for the ARP spoofing and the RKE cloning, although studies evaluate them as attacks against the PEV networks, no countermeasure has been proposed yet.

In the cyber-physical group, the fourteen (14) countermeasures proposed against power outage/overload cover most of the assets affected. However, only three (3) of them protect the data. The set of cyber-physical attacks includes countermeasures that cover most or all the affected assets and none of these attacks is without countermeasures in the literature.

The conclusions drawn regarding the open issues and the suggested future research directions are based on the current research outcomes and the security features of OCPP 2.0. In the following Sections, these open issues are described in detail.

### B. Gaps in security of architectural assets

As already mentioned, in the group of physical attacks and specifically regarding tampering, fifteen (15) countermeasures are proposed, none of which is considering the EMS protection. In the same group, EMS is one of the elements classified as a sensor attack targeted asset, with no proposed countermeasure in the literature.

In the group of cyber attacks, EMS can be targeted by MitM attacks, for which nine (9) countermeasures are proposed in the literature, while none is focusing on the EMS protection. DoS/DDoS attacks can also target EMS. However, out of the five (5) proposed DoS/DDoS countermeasures, none refers to the EMS protection. Finally, two (2) out of the two (2) proposed malware countermeasures are protecting the EMS.

In the last group of cyber-physical attacks, eight (8) out of the fourteen (14) proposed countermeasures against power outages/overloads are considering the EMS protection. Nevertheless, there is a research gap in the literature in regard to the remaining EMS cyber-physical attacks, as well as any countermeasure against them that considers EMS. These remaining attacks are: the substitution, the meter bypassing, the OTA updates tampering, the smart card cloning, the masquerading/impersonation, the FDIA, the insider attacks, and the switching attacks.

The research gap of security countermeasures for the protection of the EMS architectural asset may be justified by the fact that the OCPP function blocks, which include the EMS in the PEV network, first appeared in the latest OCPP 2.0 smart charging feature [33]. OCPP 2.0 additionally supports encryption and device management for smart charging [15]. However, as a newly included architectural asset, EMS seems to lack a strong protection against physical and cyber attacks, for which countermeasures and good practice proposals will be welcome in the future.

### C. Gaps in physical security

An OCPP-based charging implementation is vulnerable to side-channel attacks when the PLC communication is used. As already discussed in [77], the PLC circuit operates by design as an antenna, allowing the waveforms of the PLC communication to be wirelessly eavesdropped or manipulated. The main focus of these attacks are the credentials of the EV or/and the driver. These credentials are transferred from the EV to the CS via the PLC communication channel, and are then transmitted between the CS and the CSMS during the authentication process.

The proposed in the literature countermeasures for these attacks consider the generation and storage of the credentials within the EV, carried out by an embedded hardware module [37],[70]. This solution eliminates the data transmission over PLC, and, hence, it mitigates the side-channel attacks. However, the coverage of the proposed side-channel countermeasures is limited to this solution, although the PLC is widely used for the communication between the EV and the CS.

### D. Gaps in cyber security

Some of the identified attacks that affect the OCPP operations and are discussed in the relevant literature are not addressed by a countermeasure or a good practice yet. One example is the sensor attacks that target the EVs. As described in Section IV-D, these attacks affect the intra-ECU or the Vehicle-to-Everything (V2X) communications and, therefore, the OCPP operations during a compromised EV charging process. In several cases, sensor attacks are also the first step leading to FDIA, jamming or DoS attacks [96],[97]. However, the literature does not contain a work or a study regarding countermeasures specifically for this physical attack.

The ARP spoofing is a known cyber attack against the PEV network elements, and, more specifically, against the EV and the CS [60] and has similarities to the MitM attacks [71]. Nevertheless, no OCPP-specific or PEV-specific countermeasures have been proposed so far. ARP spoofing in PEV networks is currently addressed using conventional countermeasures, such as identifying an attack, relying on VPNs, static ARP, IDS implementation, and packet filtering.

This survey highlighted that the RKE cloning is another attack for which no PEV-specific or OCPP-specific countermeasures are proposed in the literature. Nonetheless, there are studies regarding the vulnerability of an EV to RKE cloning [71],[95] and, it goes without saying, that when a compromised EV enters the PEV network, the OCPP operations are consequently affected.

The gaps in the aforementioned attacks are associated with the CS and the EV. These gaps may partially be explained by the vendor-agnostic nature of OCPP. This characteristic allows the protocol to be supported by cross-vendor CSs and a variety of EV models from various manufacturers. For these reasons, OCPP security is dependent on the security of all these products. Another explanation of these gaps is the fact that OCPP 2.0 with considerations regarding security has been recently developed.

### E. Enhanced cyber-physical security

The scheduling process and the security challenges of an EV charging system led the research efforts towards the BC technology. BC is an emerging technology that can be used to secure the distribution of basic safety messages between the elements of a PEV network [162]. In the case of OCPP, the communication between an EV and the CS can be blockchain-based to ensure privacy and security for the elements identities, as well as secure metering/billing processes [114],[170]. However, BC suffers from transactions overhead, high storage requirements, increased computational time, communication overhead caused by the smart contracts interchange, and high energy consumption caused by the consensus algorithms.

As a BC enhancement, AI can be used in the transaction layer to both detect anomalies and forecast the system dynamics. AI can be used to adjust the smart contracts of the BC technology for improved EV scheduling mechanisms. In principle, the integration of AI and BC serves as the niche for a self-correcting EV charging ecosystem [106]. This can be further extended to develop new consensus algorithms that reduce the overhead in the network while considering the trade-offs in terms of security. Lastly, the implementation of the federated learning technique within the BC network would aid in providing fully-decentralised nodes [106]. These nodes are interconnected through a P2P network and, therefore, cyber security vulnerabilities are often caused by the peripheral applications, such as the digital wallets or the smart contracts [71]. Moreover, the decentralisation of energy trading may complicate the energy management. The P2P marketplaces and the local micro-grids may accelerate grid defection or may lead to severe under-utilisation of network assets [170]. Future research regarding OCPP should explore how the protocol can leverage both the BC and the AI technologies for sufficient security levels and energy management.

### F. Enhanced detection and deflection mechanisms

As described in Section V-C, the interconnection of the smart grid distributed elements introduces vectors for cyber attacks, such as the power overloading or the energy theft attacks, which can be mitigated with an ETDS. ETDSs can effectively detect energy theft attacks against AMIs and differentiate them from other cyber attacks, allowing an efficient and early detection. Despite the fact that energy is the main consumed asset of an OCPP-based charging system and the high impact of energy theft in such systems, the number of ETDSs proposed in the literature is limited [117].

## VII. Conclusion

The security of a PEV network is important for the charging service life-cycle and for all the included assets, namely the EV, the EV's driver, the charging infrastructure, and the back-end systems. The protocol used for the communication within a PEV network should also have security and privacy protecting features. OCPP is a de facto protocol for the communication between the CS, the CSMS and other back-end systems. The security of the OCPP has been studied by the protocol's developers, who introduced the OCPP 2.0 version with a set of security features and the implementation of security profiles with mandatory use of the TLS protocol. However, there still exist implementations which use older versions of the protocol with less security features.

In the literature, several studies address the security of OCPP, focusing on possible attacks against a PEV network and the protocol, and proposing a relative countermeasure. These studies have been grouped and presented here. Although they provide a wide coverage of many possible attacks against the protocol, there are some cases, such as the ARP spoofing, the RKE cloning, and the state/sensor attacks, where no valid countermeasures or good practices have been proposed. Additionally, some assets are more often spotlighted in terms of security than others. As a result, limited proposals exist in the literature regarding the security for systems such as EMS.

The blockchain technique has been the preferred method in several cases where a countermeasure was proposed. The billing process, which is an important part of the PEV network's operations, can be well-fulfilled by a blockchain-based technique, which also protects the data and other sensitive information with the use of a distributed ledger and smart contracts. The blockchain techniques are used to fill security and privacy gaps of the OCPP and this will be enhanced in the future by the combination of blockchain with AI solutions.

## REFERENCES

[1] "Open vs Closed Charging Stations: Advantages and Disadvantages," Greenlots, Tech. Rep., 2018. [Online]. Available: https://greenlots.com/wp-content/uploads/2018/10/Open-Standards-WhitePaper-compressed.pdf

[2] K. Bhargavi, N. Jayalaksmi, S. Malagi, and V. K. Jadoun, "Integration of Plug-in Electric Vehicles in Smart Grid: A Review," in *IEEE Int. Conf. on Power Electronics & IoT Applications in Renewable Energy and its Control (PARC)*, 28-29 Feb, Mathura, India, 2020, pp. 214–219.

[3] H. ElHussini, C. Assi, B. Moussa, R. Atallah, and A. Ghrayeb, "A tale of two entities: Contextualizing the security of electric vehicle charging stations on the power grid," *ACM Transactions on Internet of Things*, vol. 2, no. 2, pp. 1–21, 2021.

[4] K. Kim, J. S. Kim, S. Jeong, J. H. Park, and H. K. Kim, "Cybersecurity for autonomous vehicles: Review of attacks and defense," *Computers and Security*, vol. 103, p. 102150, 2021.

[5] "Open Charge Alliance," Open Charge Alliance (OCA), Tech. Rep., 2020. [Online]. Available: https://www.openchargealliance.org/

[6] "Enable innovation and cost efficiency with OCPP," 2022. [Online]. Available: https://www.ampeco.com/ocpp-open-charge-point-protocol/#cpo-benefits

[7] "Innovation and cost-efficiency in four letters: OCPP," 2021. [Online]. Available: http://www.current.eco/platform/ocpp

[8] "Open Charge Alliance - Our mission," 2022. [Online]. Available: https://www.openchargealliance.org/about-us/

[9] H. T. Mouftah and M. Erol-Kantarci, *Smart grid: networking, data management, and business models*. CRC Press, 2017.

[10] B. Blum, "Cyberattacks on cars increased 225% in last three years," ISRAEL21c, Tech. Rep., 2022. [Online]. Available: https://www.israel21c.org/cyberattacks-on-cars-increased-225-in-last-three-years/

[11] T. Nasr, S. Torabi, E. Bou-Harb, C. Fachkha, and C. Assi, "Power jacking your station: In-depth security analysis of electric vehicle charging station management systems," *Computers & Security*, vol. 112, p. 102511, 2022.

[12] S. Acharya, Y. Dvorkin, and R. Karri, "Public Plug-in Electric Vehicles + Grid Data: Is a New Cyberattack Vector Viable?" *IEEE Transactions on Smart Grid*, vol. 11, no. 6, pp. 5099–5113, 2020.

[13] "Open Charge Point Protocol 2.0.1," Open Charge Alliance (OCA), Tech. Rep., 2018. [Online]. Available: https://www.openchargealliance.org/protocols/ocpp-201/

[14] "Open Charge Point Protocol 1.6," Open Charge Alliance (OCA), Tech. Rep., 2015. [Online]. Available: https://www.openchargealliance.org/protocols/ocpp-16/

[15] F. Buve, M. Jansen, P. Klapwijk, and R. d. Leeuw, "OCPP 2.0.1, Part 2 - Specification," Open Charge Alliance (OCA), Tech. Rep., 2020. [Online]. Available: https://www.openchargealliance.org/protocols/ocpp-201/

[16] A. Wargers and D. Frenkel, "The world's first large-scale migration of OCPP based PEV charging infrastructure," Open Charge Alliance (OCA), Tech. Rep. [Online]. Available: https://www.openchargealliance.org/uploads/files/OCA-White_paper_on_OCPP_based_migration_version_5.0.pdf

[17] M. van Amstel, R. Ghatikar, and A. Wargers, "Importance of Open Charge Point Protocol for the Electric Vehicle Industry," Open Charge Alliance (OCA), Tech. Rep., 2016. [Online]. Available: https://www.openchargealliance.org/uploads/files/OCA-EN_whitepaper_OCPP_vs_proprietary_protocols_v1.0.pdf

[18] F. Buve, M. Jansen, and P. Klapwijk, "OCPP 2.0.1, Part 1 - Architecture & Topology," Open Charge Alliance (OCA), Tech. Rep., 2020. [Online]. Available: https://www.openchargealliance.org/protocols/ocpp-201/

[19] S. Orcioni and M. Conti, "EV smart charging with advance reservation extension to the OCPP standard," *Energies*, vol. 13, no. 12, pp. 3263–3284, 2020.

[20] J. Schlund, M. Pruckner, and R. German, "FlexAbility-Modeling and Maximizing the Bidirectional Flexibility Availability of Unidirectional Charging of Large Pools of Electric Vehicles," in *Proc. of the Eleventh ACM Int. Conf. on Future Energy Systems*, 22-26 Jun, virtual event, Australia, 2020, pp. 121–132.

[21] E. Ancillotti, R. Bruno, S. Palumbo, C. Capasso, and O. Veneri, "Experimental set-up of DC PEV charging station supported by open and interoperable communication technologies," in *IEEE Int. Symp. on Power Electronics, Electrical Drives, Automation and Motion (SPEEDAM)*, 22-24 Jun, Anacapri Capri Island, Italy, 2016, pp. 677–682.

[22] J. W. Heron, J. Jiang, H. Sun, V. Gezerlis, and T. Doukoglou, "Demand-response round-trip latency of IoT smartgrid network topologies," *IEEE Access*, vol. 6, pp. 22 930–22 937, 2018.

[23] J. W. Heron and H. Sun, "Smart electric vehicle charging with ideal and practical communications in smart grids," in *IEEE Global Communications Conf. (GLOBECOM)*, 9-13 Dec, Big Island, Hawaii, USA, 2019, pp. 1–6.

[24] L. Noel, G. Z. de Rubens, J. Kester, and B. K. Sovacool, *Vehicle-to-Grid: A Sociotechnical Transition Beyond Electric Mobility*. Springer, 2019.

[25] P. Van Aubel and E. Poll, "Security of EV-Charging Protocols," March 2021, *(in press)*. [Online]. Available: https://www.polvanaubel.com/research/chargego/protocol-security-evaluation/protocol-security-evaluation-draft-2020-03-10.pdf

[26] I. Buamod, E. Abdelmoghith, and H. T. Mouftah, "A review of OSI-based charging standards and eMobility open protocols," in *IEEE 6th Int. Conf. on the Network of the Future (NOF)*, 30 Sep-2 Oct, Montreal, Canada, 2015, pp. 9:1–9:7.

[27] D. Wellisch, J. Lenz, A. Faschingbauer, R. Pöschl, and S. Kunze, "Vehicle-to-grid AC charging station: an approach for smart charging development," *IFAC-PapersOnLine*, vol. 48, no. 4, pp. 55–60, 2015.

[28] M. Emre *et al.*, "Task 3.3.1 Review of existing power transfer solutions," EU Seventh Framework Programme, Tech. Rep., 2014.

[29] D. Wellisch, S. Kunze, and R. Pöschl, "A modular software implementation for smart charging stations," in *IEEE Int. Symp. on Smart Electric Distribution Systems and Technologies (EDST)*, 8-11 Sep, Vienna, Austria, 2015, pp. 254–259.

[30] "Background Open Charge Alliance," Open Charge Alliance (OCA), Tech. Rep., 2009. [Online]. Available: https://www.openchargealliance.org/about-us/background/

[31] "Road Vehicles - Vehicle-to-Grid Communication Interface - Part 2: Network and Application Protocol Requirements," Int. Organization for Standardization (ISO), Tech. Rep., 2014. [Online]. Available: https://www.iso.org/obp/ui/#iso:std:iso:15118:-2:ed-1:v1:en

[32] "Road vehicles — Vehicle to grid communication interface — Part 1: General information and use-case definition," Int. Organization for Standardization (ISO), Tech. Rep., 2019. [Online]. Available: https://www.iso.org/obp/ui/#iso:std:iso:15118:-1:ed-2:v1:en

[33] F. Buve, P. Klapwijk, and R. de Leeuw, "OCPP 2.0.1, Part 0 - Introduction," Open Charge Alliance (OCA), Tech. Rep., 2020. [Online]. Available: https://www.openchargealliance.org/protocols/ocpp-201/

[34] "Road Vehicles - Vehicle-to-Grid Communication Interface - Part 3: Physical and data link layer requirements," Int. Organization for Standardization (ISO), Tech. Rep., 2015. [Online]. Available: https://www.iso.org/obp/ui/#iso:std:iso:15118:-3:en

[35] M. Parchomiuk, A. Moradewicz, and H. Gawiński, "An Overview of Electric Vehicles Fast Charging Infrastructure," in *IEEE Progress in Applied Electrical Engineering (PAEE)*, 17-21 Jun, Koscielisko, Poland, 2019, pp. 26:1–26:5.

[36] C. Hodge *et al.*, "Vehicle Cybersecurity Threats and Mitigation Approaches," National Renewable Energy Laboratory (NREL), Tech. Rep. NREL/TP-5400-74247, 2019. [Online]. Available: www.nrel.gov/publications.%0Ahttps://www.nrel.gov/docs/fy19osti/74247.pdf

[37] L. Buschlinger, M. Springer, and M. Zhdanova, "Plug-and-patch: Secure value added services for electric vehicle charging," in *ACM Proc. of the 14th Int. Conf. on Availability, Reliability and Security*, 26-29 Aug, Canterbury, UK, 2019, pp. 2:1–2:10.

[38] Z. Jakó and Á. Knapp, "Business Scenarios and Data Flow in NeMo Hyper-Network," in *IEEE Int. Conf. on Smart Systems and Technologies (SST)*, 10-12 Oct, Osijek, Croatia, 2018, pp. 139–144.

[39] J. Antoun, M. E. Kabir, B. Moussa, R. Atallah, and C. Assi, "A Detailed Security Assessment of the EV Charging Ecosystem," *IEEE Network*, vol. 34, no. 3, pp. 200–207, 2020.

[40] R. Bilolikar, A. Magal, N. Lepre, and J. Korsh, "Scaling Up Electric Vehicle Charging Infrastructure," Natural Resources Defense Council (NRDC), Tech. Rep., 2020. [Online]. Available: https://www.nrdc.org/sites/default/files/charging-infrastructure-best-parctices-202007.pdf

[41] B. Douglas, J. McDonald, N. DeForest, and C. Gehbauer, "Los Angeles Air Force Base Vehicle-to-Grid Demonstration: Final Project Report," California Energy Commission, Tech. Rep. CEC-500-2018-025, 2017.

[42] "ECHONET Lite Specification, Version 1.13," ECHONET Consortium, Tech. Rep., 2018. [Online]. Available: https://echonet.jp/spec_v113_lite_en/

[43] U. Willrett, "Grid integration e-mobility–Developments and challenges," in *19. Internationales Stuttgarter Symp.* 19-20 Mar, Stuttgart, Germany: Springer, 2019, pp. 320–330.

[44] V. C. Pham, Y. Makino, K. Pho, Y. Lim, and Y. Tan, "IoT Area Network Simulator For Network Dataset Generation," *J. of Information Processing*, vol. 28, pp. 668–678, 2020.

[45] U. Willrett, "Standards for Implementing Smart Charging," *ATZ worldwide*, vol. 122, no. 12, pp. 64–67, 2020.

[46] M. Erol-Kantarci and H. T. Mouftah, "Pervasive Energy Management for the Smart Grid: Towards a Low Carbon Economy," in *Pervasive Communications Handbook*. CRC Press, 2017, pp. 251–269.

[47] "Open Intercharge Protocol," Hubject GMBH, Tech. Rep., 2018. [Online]. Available: https://www.hubject.com/en/downloads/oicp/

[48] "Road vehicles — Vehicle to grid communication interface — Part 4: Network and application protocol conformance test," Int. Organization for Standardization (ISO), Tech. Rep., 2018. [Online]. Available: https://www.iso.org/obp/ui/#iso:std:iso:15118:-4:ed-1:v1:en

[49] "Road vehicles — Vehicle to grid communication interface — Part 5: Physical layer and data link layer conformance test," Int. Organization for Standardization (ISO), Tech. Rep., 2018. [Online]. Available: https://www.iso.org/obp/ui/#iso:std:iso:15118:-5:ed-1:v2:en

[50] "Road vehicles – Vehicle to grid communication interface – Part 8: Physical layer and data link layer requirements for wireless communication," Int. Organization for Standardization (ISO), Tech. Rep., 2020. [Online]. Available: https://www.iso.org/obp/ui/#iso:std:iso:15118:-8:ed-2:v1:en

[51] "Road vehicles — Vehicle to grid communication interface — Part 9: Physical and data link layer conformance test for wireless communication," Int. Organization for Standardization (ISO), Tech. Rep., 2021. [Online]. Available: https://www.iso.org/obp/ui/#iso:std:iso:15118:-9:dis:ed-1:v1:en

[52] U. Willrett, "Electric vehicles–enablers for the energy transition?" in *Netzintegration der Elektromobilität 2018*. Springer, 2018, pp. 56–65.

[53] M. Neaimeh and P. B. Andersen, "Mind the gap-open communication protocols for vehicle grid integration," *Energy Informatics*, vol. 3, no. 1, pp. 1:1–1:17, 2020.

[54] R. Metere, M. Neaimeh, C. Morisset, C. Maple, X. Bellekens, and R. M. Czekster, "Securing the Electric Vehicle Charging Infrastructure," *arXiv preprint arXiv:2105.02905*, p. 39, 2021.

[55] A. Gopstein, C. Nguyen, C. O'Fallon, D. Wollman, and N. Hasting, "NIST Framework and Roadmap for Smart Grid Interoperability Standards, Release 4.0," NIST Special Publication 800-53, Tech. Rep., 2020.

[56] D. G. Márquez, C. M. Fernández, P. R. Pinos, and K. Piotrowski, "Networking layer specification," EU Horizon 2020, Tech. Rep., 2020.

[57] I. S. F. Gomes, Y. Perez, and E. Suomalainen, "Coupling small batteries and PV generation: a review," *Renewable and Sustainable Energy Reviews*, vol. 126, p. 109835, 2020.

[58] R. Rodríguez *et al.*, "ICT requirements specifications," EU Horizon 2020, Tech. Rep., 2016.

[59] A. Sundararajan, A. Chavan, D. Saleem, and A. I. Sarwat, "A Survey of Protocol-Level Challenges and Solutions for Distributed Energy Resource Cyber-Physical Security," *Energies*, vol. 11, no. 9, pp. 2360–2381, 2018.

[60] Y.-W. Chung, *Electric Vehicle–Smart Grid Integration: Load Modeling, Scheduling, and Cyber Security*. University of California, Los Angeles, 2020.

[61] M. Marinelli *et al.*, "Electric Vehicles Demonstration Projects-An Overview Across Europe," in *IEEE 55th Int. Universities Power Engineering Conf. (UPEC)*, 1-4 Sep, Turin, Italy, 2020, pp. 19:1–19:6.

[62] W. Han and Y. Xiao, "Privacy preservation for V2G networks in smart grid: A survey," *Computer Communications*, vol. 91-92, pp. 17–28, 2016.

[63] C. Bernardini, M. R. Asghar, and B. Crispo, "Security and privacy in vehicular communications: Challenges and opportunities," *Vehicular Communications*, vol. 10, pp. 13–28, 2017.

[64] C. Alcaraz, J. Lopez, and S. Wolthusen, "OCPP Protocol: Security Threats and Challenges," *IEEE Transactions on Smart Grid*, vol. 8, no. 5, pp. 2452–2459, 2017.

[65] Z. Pourmirza and S. Walker, "Electric Vehicle Charging Station: Cyber Security Challenges and Perspective," in *IEEE 9th Int. Conf. on Smart Energy Grid Engineering (SEGE)*, 11-13 Aug, Oshawa, Canada, 2021, pp. 111–116.

[66] M. S. Raboaca *et al.*, "An overview and performance evaluation of open charge point protocol from an electromobility concept perspective," *Int. J. of Energy Research*, pp. 1–21, 2021.

[67] C. Levy-Bencheton, E. Darra, D. Bachlechner, and M. Friedewald, "Cyber security for smart cities—An architecture model for public transport," European Union Agency for Network and Information Security (ENISA), Tech. Rep., 2015.

[68] H. van den Brink, "EV charging Systems Security Requirements," European Network for Cyber Security (ENCS), Tech. Rep., 2017.

[69] Y. Mo *et al.*, "Cyber-physical security of a smart grid infrastructure," *Proc. of the IEEE*, vol. 100, no. 1, pp. 195–209, 2012.

[70] A. Fuchs, D. Kern, C. Krauß, and M. Zhdanova, "HIP: HSM-based identities for plug-and-charge," in *ACM Proc. of the 15th Int. Conf. on Availability, Reliability and Security*, 25-28 Aug, Dublin, Ireland, 2020, pp. 33:1–33:6.

[71] Z. El-Rewini, K. Sadatsharan, D. F. Selvaraj, S. J. Plathottam, and P. Ranganathan, "Cybersecurity challenges in vehicular communications," *Vehicular Communications*, vol. 23, pp. 100 214–100 242, 2020.

[72] H. van den Brink, "The need for cybersecurity within the electric vehicle infrastructure - A study on the use of digital signatures in the electric vehicle infrastructure," in *30th Int. Electric Vehicle Symp. & Exhibition, EVS30*. 9-11 Oct, Stuttgart, Germany: European Association for Electromobility (AVERE), 2017, pp. 7:5:1–7:5:10.

[73] M. van Eekelen, E. Poll, E. Hubbers, B. Vieira, and F. van den Broek, "An end-to-end security design for smart EV-charging for Enexis and ElaadNL," Technische Universiteit Eindhoven and Radboud Universiteit Nijmegen, Tech. Rep., 2014.

[74] M. Mustafa, N. Zhang, G. Kalogridis, and Z. Fan, "Roaming electric vehicle charging and billing: an anonymous multi-user protocol," in *Proc. of the IEEE Int. Conf. on Smart Grid Communications (SmartGridComm)*, 3-6 Nov, Venice, Italy, 2014, pp. 939–945.

[75] A. C.-F. Chan and J. Zhou, "Cyber–physical device authentication for the smart grid electric vehicle ecosystem," *IEEE J. on Selected Areas in Communications*, vol. 32, no. 7, pp. 1509–1517, 2014.

[76] A. A. Soares, D. M. Mattos, Y. Lopes, D. S. Medeiros, N. C. Fernandes, and D. C. Muchaluat-Saade, "An Efficient Authentication Mechanism based on Software-Defined Networks for Electric Vehicles," in *IEEE 28th Int. Symp. on Industrial Electronics (ISIE)*, 12-14 Jun, Vancouver, Canada, 2019, pp. 2471–2476.

[77] R. Baker and I. Martinovic, "Losing the car keys: Wireless phy-layer insecurity in EV charging," in *Proc. of the 28th USENIX Security Symp.*, 14-16 Aug, Santa Clara, CA, USA, 2019, pp. 407–424.

[78] M. Khodari, A. Rawat, M. Asplund, and A. Gurtov, "Decentralized firmware attestation for in-vehicle networks," in *CPSS 2019 - Proc. of the 5th ACM Cyber-Physical System Security Workshop*, 8 Jul, Auckland, Australia, 2019, pp. 47–56.

[79] A. Rawat, M. Khodari, M. Asplund, and A. Gurtov, "Decentralized Firmware Attestation for In-Vehicle Networks," *ACM Trans. Cyber-Phys. Syst*, vol. 5, no. 1, p. 23, 2020.

[80] D. Zelle, M. Springer, M. Zhdanova, and C. Krauß, "Anonymous charging and billing of electric vehicles," in *ACM Proc. of the 13th Int. Conf. on Availability, Reliability and Security*, 27-30 Aug, Hamburg, Germany, 2018, pp. 22:1–22:10.

[81] S. M. Danish, K. Zhang, H. A. Jacobsen, N. Ashraf, and H. K. Qureshi, "BlockEV: Efficient and Secure Charging Station Selection for Electric Vehicles," *IEEE Transactions on Intelligent Transportation Systems*, vol. 22, no. 7, pp. 4194–4211, 2020.

[82] D. Gabay, K. Akkaya, and M. Cebe, "Privacy-preserving Authentication scheme for Connected Electric Vehicles Using Blockchain and Zero Knowledge Proofs," *IEEE Transactions on Vehicular Technology*, vol. 69, no. 6, pp. 5760–5772, 2020.

[83] H. Nicanfar, S. Hosseininezhad, P. TalebiFard, and V. C. Leung, "Robust privacy-preserving authentication scheme for communication

between Electric Vehicle as Power Energy Storage and power stations," in *IEEE Conf. on Computer Communications Workshops (INFOCOM WKSHPS)*, 14-19 Apr, Turin, Italy, 2013, pp. 55–60.

[84] H. Li, G. Dán, and K. Nahrstedt, "Portunes+: Privacy-preserving fast authentication for dynamic electric vehicle charging," *IEEE Transactions on Smart Grid*, vol. 8, no. 5, pp. 2305–2313, 2016.

[85] P. Gope and B. Sikdar, "An efficient privacy-preserving authentication scheme for energy internet-based vehicle-to-grid communication," *IEEE Transactions on Smart Grid*, vol. 10, no. 6, pp. 6607–6618, 2019.

[86] A. Irshad, M. Usman, S. A. Chaudhry, H. Naqvi, and M. Shafiq, "A Provably Secure and Efficient Authenticated Key Agreement Scheme for Energy Internet-Based Vehicle-to-Grid Technology Framework," *IEEE Transactions on Industry Applications*, vol. 56, no. 4, pp. 4425–4435, 2020.

[87] G. Bansal, N. Naren, V. Chamola, B. Sikdar, N. Kumar, and M. Guizani, "Lightweight Mutual Authentication Protocol for V2G Using Physical Unclonable Function," *IEEE Transactions on Vehicular Technology*, vol. 69, no. 7, pp. 7234–7246, 2020.

[88] "Trusted Platform Module Library 2.0, Revision 01.59," Trusted Computing Group (TCG), Tech. Rep., 2019. [Online]. Available: https://trustedcomputinggroup.org/resource/tpm-library-specification/

[89] A. Fuchs, D. Kern, C. Krauss, and M. Zhdanova, "TrustEV: trustworthy electric vehicle charging and billing," in *Proc. of the 35th Annual ACM Symp. on Applied Computing*, 30 Mar-3 Apr, Brno, Czech Republic, 2020, pp. 1706–1715.

[90] T. Lipman *et al.*, "Open-Source, Open-Architecture Software Platform for Plug-In Electric Vehicle Smart Charging in California," University of California – Berkeley, Transportation Sustainability Research Center, Tech. Rep., 2020.

[91] S. Lightman and T. Brewer, *Symposium on federally funded research on cybersecurity of electric vehicle supply equipment (EVSE)*. US Department of Commerce, National Institute of Standards and Technology (NIST), 2020.

[92] A. Heinrich and R. Heddergott, "Secure and User-Friendly EV Charging A Comparison of Autocharge and ISO 15118's Plug & Charge," V2G Clarity, Hubject, Tech. Rep., 2019. [Online]. Available: {https://uploads-ssl.webflow.com/607417b42ba2bfea543956dd/60c33c6cc823f9bb6bd4f275_Whitepaper-Autocharge-vs-ISO15118-Plug-and-Charge.pdf}

[93] A. R. Short, H. C. Leligou, and E. Theocharis, "Execution of a Federated Learning process within a smart contract," in *IEEE Int. Conf. on Consumer Electronics (ICCE)*, 10-12 Jan, Las Vegas, NV, USA, 2021, pp. 1–4.

[94] G. Dhanush, K. S. Raj, and P. Kumar, "Blockchain Aided Predictive Time Series Analysis in Supply Chain System," in *Innovations in Electrical and Electronic Engineering*. Springer, 2021, pp. 913–925.

[95] I. Stellios, P. Kotzanikolaou, M. Psarakis, C. Alcaraz, and J. Lopez, "A survey of iot-enabled cyberattacks: Assessing attack paths to critical infrastructures and services," *IEEE Communications Surveys & Tutorials*, vol. 20, no. 4, pp. 3453–3495, 2018.

[96] X. Luo, Y. Li, X. Wang, and X. Guan, "Interval Observer-Based Detection and Localization against False Data Injection Attack in Smart Grids," *IEEE Internet of Things J.*, vol. 8, no. 2, pp. 657–671, 2021.

[97] Y. Fraiji, L. B. Azzouz, W. Trojet, and L. A. Saidane, "Cyber security issues of Internet of Electric Vehicles," in *IEEE Wireless Communications and Networking Conf. (WCNC)*, 15-18 Apr, Barcelona, Spain, 2018, pp. 1–6.

[98] A. Irshad, M. Usman, S. A. Chaudhry, H. Naqvi, and M. Shafiq, "A provably secure and efficient authenticated key agreement scheme for Energy Internet based Vehicle-to-Grid technology framework," *IEEE Transactions on Industry Applications*, vol. 56, no. 4, pp. 4425–4435, 2020.

[99] M. E. Kabir, M. Ghafouri, B. Moussa, and C. Assi, "A Two-Stage Protection Method for Detection and Mitigation of Coordinated EVSE Switching Attacks," *IEEE Transactions on Smart Grid*, vol. 12, no. 5, pp. 4377–4388, 2021.

[100] J. E. Rubio, C. Alcaraz, and J. Lopez, "Addressing Security in OCPP: Protection Against Man-in-The-Middle Attacks," in *IEEE 9th IFIP Int. Conf. on New Technologies, Mobility and Security (NTMS)*, vol. 2018-January, 26-28 Feb, Paris, France, 2018, pp. 6:1–6:5.

[101] B. Vaidya and H. T. Mouftah, "Multimodal and Multi-pass Authentication Mechanisms for Electric Vehicle Charging Networks," in *IEEE Int. Wireless Communications and Mobile Computing, IWCMC 2020*, 15-19 Jun, Limassol, Cyprus, 2020, pp. 371–376.

[102] Y. Li and B. Hu, "A Consortium Blockchain-Enabled Secure and Privacy-Preserving Optimized Charging and Discharging Trading Scheme for Electric Vehicles," *IEEE Transactions on Industrial Informatics*, vol. 17, no. 3, pp. 1968–1977, 2020.

[103] ——, "An iterative two-layer optimization charging and discharging trading scheme for electric vehicle using consortium blockchain," *IEEE Transactions on Smart Grid*, vol. 11, no. 3, pp. 2627–2637, 2019.

[104] W. Choi, H. J. Jo, S. Woo, J. Y. Chun, J. Park, and D. H. Lee, "Identifying ECUs Using Inimitable Characteristics of Signals in Controller Area Networks," *IEEE Transactions on Vehicular Technology*, vol. 67, no. 6, pp. 4757–4770, 2018.

[105] W. Choi, K. Joo, H. J. Jo, M. C. Park, and D. H. Lee, "VoltageIDS: Low-Level Communication Characteristics for Automotive Intrusion Detection System," *IEEE Transactions on Information Forensics and Security*, vol. 13, no. 8, pp. 2114–2129, 2018.

[106] H. ElHusseini, C. Assi, B. Moussa, R. Attallah, and A. Ghrayeb, "Blockchain, AI and Smart Grids: The Three Musketeers to a Decentralized EV Charging Infrastructure," *IEEE Internet of Things Magazine*, vol. 3, no. 2, pp. 24–29, 2020.

[107] M. Conti, D. Donadel, R. Poovendran, and F. Turrin, "EVExchange: A Relay Attack on Electric Vehicle Charging System," *arXiv preprint arXiv:2203.05266*, p. 20, 2022.

[108] Y. Li and B. Hu, "An Iterative Two-Layer Optimization Charging and Discharging Trading Scheme for Electric Vehicle Using Consortium Blockchain," *IEEE Transactions on Smart Grid*, vol. 11, no. 3, pp. 2627–2637, 2020.

[109] A. G. Morosan and F. Pop, "OCPP security - Neural network for detecting malicious traffic," in *ACM Proc. of the 2017 Research in Adaptive and Convergent Systems, RACS 2017*, vol. 2017-January, 13-16 Oct, Gwangju , Republic of Korea, 2017, pp. 190–195.

[110] F. Knirsch, A. Unterweger, and D. Engel, "Privacy-preserving blockchain-based electric vehicle charging with dynamic tariff decisions," *Computer Science - Research and Development*, vol. 33, pp. 71–79, 2017.

[111] J. Li *et al.*, "Decentralized On-Demand Energy Supply for Blockchain in Internet of Things: A Microgrids Approach," *IEEE Transactions on Computational Social Systems*, vol. 6, no. 6, pp. 1395–1406, 2019.

[112] S. M. Danish, K. Zhang, and H. A. Jacobsen, "A Blockchain-Based Privacy-Preserving Intelligent Charging Station Selection for Electric Vehicles," in *IEEE Int. Conf. on Blockchain and Cryptocurrency (ICBC)*, 2-6 May, Toronto, ON, Canada, 2020, pp. PD2A6:1–PD2A6:3.

[113] A. Gharaibeh *et al.*, "Smart cities: A survey on data management, security, and enabling technologies," *IEEE Communications Surveys & Tutorials*, vol. 19, no. 4, pp. 2456–2501, 2017.

[114] P. van Aubel, E. Poll, and J. Rijneveld, "Non-Repudiation and End-to-End Security for Electric-Vehicle Charging," in *IEEE PES Innovative Smart Grid Technologies Europe (ISGT-Europe)*, 29 Sep-2 Oct, Bucharest, Romania, 2019, pp. 191–195.

[115] A. Ghasempour and J. H. Gunther, "Finding the optimal number of aggregators in machine-to-machine advanced metering infrastructure architecture of smart grid based on cost, delay, and energy consumption," in *13th IEEE Annual Consumer Communications Networking Conf. (CCNC)*, 9-12 Jan, Las Vegas, NV, USA, 2016, pp. 960–963.

[116] A. A. Korba, N. Tamani, Y. Ghamri-Doudane, and N. E. I. Karabadji, "Anomaly-based framework for detecting power overloading cyberattacks in smart grid AMI," *Computers & Security*, vol. 96, p. 101896, 2020.

[117] P. Jokar, N. Arianpoo, and V. C. M. Leung, "Electricity Theft Detection in AMI Using Customers' Consumption Patterns," *IEEE Transactions on Smart Grid*, vol. 7, no. 1, pp. 216–226, 2016.

[118] S. Acharya, Y. Dvorkin, and R. Karri, "Public plug-in electric vehicles+ grid data: Is a new cyberattack vector viable?" *IEEE Transactions on Smart Grid*, vol. 11, no. 6, pp. 5099–5113, 2020.

[119] R. Uhlig, M. Stoetzel, M. Stiegler, S. Lamberth, and A. Kubis, "Hybrid Cascaded Operation of Distribution Grids," in *Int. ETG-Congress 2019; ETG Symp.* 8-9 May, Esslingen, Germany: VDE, 2019, pp. 7:1–7:6.

[120] A. Kubis, M. Boller, J. Kemper, R. Uhlig, M. Stötzel, and M. Stiegler, "Enhancing operational awareness of distribution system operators with a semi-autonomous intelligent grid control suite," in *CIRED 25th Int. Conf. on Electricity Distribution, CIRED 2019*. 3-6 Jun, Madrid, Spain: AIM, 2019, pp. 1850:1–1850:5.

[121] M. H. Amini, J. Mohammadi, and S. Kar, "Distributed holistic framework for smart city infrastructures: tale of interdependent electrified transportation network and power grid," *IEEE Access*, vol. 7, pp. 157 535–157 554, 2019.

[122] G. Putrus *et al.*, "Overview SEEV4-city playing field state-of-the-art assessment of smart charging and vehicle 2 Grid services," Amsterdam University of Applied Sciences, Tech. Rep., 2020.

[123] S. Übermasser *et al.*, "Optimized and enhanced grid architecture for electric vehicles in Europe," *Elektrotech. Inftech.*, vol. 134, no. 1, pp. 78–85, 2017.

[124] J. Kirby and F. Hassan, "AC Recharging Infrastructure for EVs and future smart grids - A review," *IEEE Proc. of the Universities Power Engineering Conf.*, pp. 1–6, 2012.

[125] B. Zhang, R. B. Carlson, J. G. Smart, E. J. Dufek, and B. Liaw, "Challenges of future high power wireless power transfer for light-duty electric vehicles—-technology and risk management," *eTransportation*, vol. 2, p. 100012, 2019.

[126] G. Li, J. Wu, J. Li, T. Ye, and R. Morello, "Battery status sensing software-defined multicast for V2G regulation in smart grid," *IEEE Sensors J.*, vol. 17, no. 23, pp. 7838–7848, 2017.

[127] T. Kobashi *et al.*, "Chapter 9 - Smart city and ICT infrastructure with vehicle to X applications toward urban decarbonization," in *Urban Systems Design*. Elsevier, 2020, pp. 289–333.

[128] M. Nürnberg and S. Iwan, "Application of Telematics Solutions for Improvement the Availability of Electric Vehicles Charging Stations," in *Development of Transport by Telematics*, vol. 1049. Springer, 2019, pp. 287–301.

[129] M. Aspragkathos, *A systems approach to designing new mobility and smart grid services: the World's Smartest Grid showcase*. Technische Universiteit Eindhoven, 2014.

[130] V. Delgado-Gomes *et al.*, "H2020-646184 NOBEL GRID New Cost Efficient Business Models for Flexible Smart Grids," Athens University of Economics and Business (AUEB), Tech. Rep., 2016. [Online]. Available: http://stecon.cs.aueb.gr/media/1203/nobel-grid-d26-final-nobelgrid-business-models.pdf

[131] Y. Zhang, R. Nakanishi, M. Sasabe, and S. Kasahara, "Combining IOTA and Attribute-Based Encryption for Access Control in the Internet of Things," *Sensors*, vol. 21, no. 15:5053, 2021.

[132] H. Liu, Y. Zhang, and T. Yang, "Blockchain-enabled security in electric vehicles cloud and edge computing," *IEEE Network*, vol. 32, no. 3, pp. 78–83, 2018.

[133] G. Xu, P. Moulema, L. Ge, H. Song, and W. Yu, "A Unified Framework for Secured Energy Resource Management," *Smart Grid: Networking, Data Management, and Business Models*, pp. 73–96, 2016.

[134] A. Iqbal, A. A. Khan, V. Kumar, and M. Ahmad, "A Mutual Authentication and Key Agreement Protocol for Vehicle to Grid Technology," in *Innovations in Electrical and Electronic Engineering*. Springer, 2021, vol. 756, pp. 863–875.

[135] J. Kester, L. Noel, X. Lin, G. Zarazua de Rubens, and B. K. Sovacool, "The coproduction of electric mobility: Selectivity, conformity and fragmentation in the sociotechnical acceptance of vehicle-to-grid (V2G) standards," *J. of Cleaner Production*, vol. 207, pp. 400–410, 2019.

[136] K. C. Sou, H. Sandberg, and K. H. Johansson, "Electric power network security analysis via minimum cut relaxation," in *50th IEEE Conf. on Decision and Control and European Control Conf.*, 12-15 Dec, Orlando, FL, USA, 2011, pp. 4054–4059.

[137] M. N. Kurt, Y. Yilmaz, and X. Wang, "Distributed Quickest Detection of Cyber-Attacks in Smart Grid," *IEEE Transactions on Information Forensics and Security*, vol. 13, pp. 2015–2030, 2018.

[138] D. Hall and N. Lutsey, "Literature review on power utility best practices regarding electric vehicles," Int. Council on Clean Transportation (ICCT), Tech. Rep., 2017.

[139] F. Wu, J. Gibbs, S. Kleinbaum, and C. Schutte, "FY2017 Materials Annual Progress Report," US Dept. of Energy (DOE), Washington DC (United States), Tech. Rep., 2018.

[140] D. Wohlschlager, S. Haas, and A. Neitz-Regett, "Comparative environmental impact assessment of ICT for smart charging of electric vehicles in Germany," *Procedia CIRP*, vol. 105, pp. 583–588, 2022.

[141] J. Howden, L. Maglaras, and M. A. Ferrag, "The security aspects of automotive over-the-air updates," *Int. J. of Cyber Warfare and Terrorism (IJCWT)*, vol. 10, no. 2, pp. 64–81, 2020.

[142] J. Díaz *et al.*, "Electric vehicle charging points mobile application," Google Patents, Tech. Rep., Jul 2018, US Patent App. 15/849,811.

[143] U. Satapathy, B. K. Mohanta, D. Jena, and S. Sobhanayak, "An ECC based Lightweight Authentication Protocol for Mobile Phone in Smart Home," in *IEEE 13th Int. Conf. on industrial and information systems (ICIIS)*, 1-2 Dec, Rupnagar, India, 2018, pp. 303–308.

[144] R. B. Bobba, K. M. Rogers, Q. Wang, K. N. Himanshu Khurana, and T. J. Overbye, "Detecting False Data Injection Attacks on DC State Estimation," in *1st Workshop on Secure Control Systems (SCS 2010), CPSWEEK2010*, 12 Apr, Stockholm, Sweden, 2010, pp. 1–9.

[145] D. Chaum, "Blind signatures for untraceable payments," *Advances in Cryptology*, pp. 199–203, 1983.

[146] H. Wang, B. Qin, Q. Wu, L. Xu, and J. Domingo-Ferrer, "TPP: Traceable privacy-p- reserving communication and precise reward for vehicle-to-grid networks in smart grids," *IEEE Trans. Inf. Forensics Secur.*, vol. 10, no. 11, pp. 2340–2346, 2015.

[147] M. Au, J. Liu, J. Fang, Z. Jiang, W. Susilo, and J. Zhou, "A new payment system for enhancing location privacy of electric vehicles," *IEEE Transactions on Vehicular Technology*, vol. 63, no. 1, pp. 3–18, 2014.

[148] A. Agarwal and R. Saraswat, "A survey of group signature technique, its applications and attacks," *Int. J. of Engineering and Innovative Technology (IJEIT)*, vol. 2, no. 10, pp. 28–35, 2013.

[149] H. H. Liu, Ning, Y. Zhang, and M. Guizani, "Battery status-aware authentication scheme for V2G networks in smart grid," *IEEE Transactions on Smart Grid*, vol. 4, no. 1, pp. 99–110, 2013.

[150] H. Liu, H. Ning, Y. Zhang, Q. Xiong, and L. Yan, "Role-dependent privacy preservation for secure V2G networks in the smart grid," *IEEE Transactions on Information Forensics and Security*, vol. 9, no. 2, pp. 208–220, 2014.

[151] C. Rottondi, S. Fontana, and G. Verticale, "Enabling privacy in vehicle-to-grid interactions for battery recharging," *Energies*, vol. 7, no. 5, pp. 2780–2798, 2014.

[152] S. Ozdemir and Y. Xiao, "Integrity protecting hierarchical concealed data aggregation for wireless sensor networks," *Computer Networks*, vol. 55, no. 8, pp. 1735–1746, 2011.

[153] Z. Wang and G. Zheng, "Residential Appliances Identification and Monitoring by a Nonintrusive Method," *IEEE Transactions on Smart Grid*, vol. 3, no. 1, pp. 80–92, 2012.

[154] R. Lu, X. Liang, X. Li, X. Lin, and X. Shen, "EPPA: An Efficient and Privacy-Preserving Aggregation Scheme for Secure Smart Grid Communications," *IEEE Transactions on Parallel and Distributed Systems*, vol. 23, no. 9, pp. 1621–1631, 2012.

[155] H. Chen, Y. Xiao, X. Hong, F. Hu, and J. Xi, "A survey of anonymity in wireless communication systems," *Security and Communication Networks*, vol. 2, no. 5, pp. 427–444, 2009.

[156] N. Saxena and B. J. Choi, "Authentication Scheme for Flexible Charging and Discharging of Mobile Vehicles in the V2G Networks," *IEEE Transactions on Information Forensics and Security*, vol. 11, no. 7, pp. 1438–1452, 2016.

[157] S. Lee, Y. Park, H. Lim, and T. Shon, "Study on analysis of security vulnerabilities and countermeasures in ISO/IEC 15118 based electric vehicle charging technology," in *IEEE Proc. of Int. Conf. on IT Convergence and Security (ICITCS)*, 28-30 Oct, Beijing, China, 2018, pp. T8P1:1–T8P1:4.

[158] K. Bao, H. Valev, M. Wagner, and H. Schmeck, "A threat analysis of the vehicle-to-grid charging protocol ISO 15118," *Computer Science - Research and Development*, vol. 33, pp. 3–12, 2018.

[159] A. A. Cárdenas *et al.*, "A Framework for Evaluating Intrusion Detection Architectures in Advanced Metering Infrastructures," *IEEE Transactions on Smart Grid*, vol. 5, no. 2, pp. 906–915, 2014.

[160] B. Genge, P. Haller, C. Dumitru, and C. Enăchescu, "Designing Optimal and Resilient Intrusion Detection Architectures for Smart Grids," *IEEE Transactions on Smart Grid*, vol. 8, no. 5, pp. 2440–2451, 2017.

[161] K. Wang, M. Du, S. Maharjan, and Y. Sun, "Strategic Honeypot Game Model for Distributed Denial of Service Attacks in the Smart Grid," *IEEE Transactions on Smart Grid*, vol. 8, no. 5, pp. 2474–2482, 2017.

[162] D. Kosmanos *et al.*, "A novel Intrusion Detection System against Spoofing Attacks in Connected Electric Vehicles," *Array*, vol. 5, p. 100013, 2020.

[163] K. Gai, M. Qiu, Z. Ming, H. Zhao, and L. Qiu, "Spoofing-Jamming Attack Strategy Using Optimal Power Distributions in Wireless Smart Grid Networks," *IEEE Transactions on Smart Grid*, vol. 8, no. 5, pp. 2431–2439, 2017.

[164] M. Mustafa, N. Zhang, G. Kalogridis, and Z. Fan, "Smart electric vehicle charging: Security Analysis," in *Proc. of the IEEE PES Innovative Smart Grid Technologies (ISGT)*, 24-27 Feb, Washington, DC, USA, 2013, pp. 46:1–46:6.

[165] J. Liu, M. Au, W. Susilo, and J. Zhou, "Enhancing location privacy for electric vehicles (at the right time)," in *Proc. of the 17th European Symp. on Research in Computer Security*. 10-12 Sep, Pisa, Italy: Springer, 2012, pp. 397–414.

[166] Y. Zhang, S. Gjessing, H. Liu, H. Ning, L. Yang, and M. Guizani, "Securing vehicle–to-grid communications in the smart grid," *IEEE Wireless Communications*, vol. 20, no. 6, pp. 66–73, 2013.

[167] Y. Zhang, Y. Xiao, K. Ghaboosi, J. Zhang, and H. Deng, "A survey of cyber crimes," *Security and Communication Networks*, vol. 5, no. 4, pp. 422–437, 2012.

[168] J. Chen, Y. Zhang, and W. Su, "An anonymous authentication scheme for plug-in electric vehicles joining to charging/discharging station in vehicle-to-Grid (V2G) networks," *IEEE China Comm.*, vol. 12, no. 3, pp. 9–19, 2015.

[169] A. C.-F. Chan and J. Zhou, "Cyber–Physical Device Authentication for the Smart Grid Electric Vehicle Ecosystem," *IEEE J. on Selected Areas in Communications*, vol. 32, no. 7, pp. 1509–1517, 2014.

[170] M. Andoni *et al.*, "Blockchain technology in the energy sector: A systematic review of challenges and opportunities," *Renewable and Sustainable Energy Reviews*, vol. 100, pp. 143–174, 2019.